\documentclass[aps,prx,amsmath,amssymb,amsfonts,superscriptaddress,notitlepage,reprint,longbibliography,floatfix
]{revtex4-2}

\usepackage{graphicx}
\usepackage{dcolumn}
\usepackage{bm}
\usepackage{amssymb}
\usepackage{amsmath}
\usepackage[svgnames]{xcolor}
\usepackage[english]{babel}
\usepackage{braket}
\usepackage{microtype}
\usepackage{bbm}
\usepackage{tikz}
\usepackage{siunitx}
\usepackage{physics}
\usepackage{csquotes}
\usepackage{multirow}
\usepackage{comment}
\usepackage[T1]{fontenc}

\usepackage{float}
\makeatletter
\let\newfloat\newfloat@ltx
\makeatother
\usepackage{algorithm}
\usepackage{algpseudocode}

\usepackage[capitalise]{cleveref}

\begin{document}

\title{Individual solid-state nuclear spin qubits with coherence exceeding seconds}

\author{James~O'Sullivan}
\thanks{Both authors contributed equally to this work}
\author{Jaime~Travesedo}
\thanks{Both authors contributed equally to this work}
\author{Louis~Pallegoix}
\author{Zhiyuan~W.~Huang}

\affiliation{Quantronics group, Service de Physique de l'\'Etat Condens\'e  (CNRS, UMR\ 3680),\\IRAMIS, CEA-Saclay, Universit\'e Paris-Saclay, 91191 Gif-sur-Yvette, France}

\author{Patrick~Hogan}
\affiliation{London Centre for Nanotechnology, UCL, 17-19 Gordon Street, London, WC1H 0AH, UK}

\author{Alexandre~S.~May}


\affiliation{Quantronics group, Service de Physique de l'\'Etat Condens\'e  (CNRS, UMR\ 3680),\\IRAMIS, CEA-Saclay, Universit\'e Paris-Saclay, 91191 Gif-sur-Yvette, France}
\affiliation{Alice$\&$Bob, 53 boulevard du Général Martial Valin, 75015 Paris}

\author{Boris~Yavkin}
\affiliation{Quantronics group, Service de Physique de l'\'Etat Condens\'e  (CNRS, UMR\ 3680),\\IRAMIS, CEA-Saclay, Universit\'e Paris-Saclay, 91191 Gif-sur-Yvette, France}

\author{Sen~Lin}
\author{Renbao~Liu}
\affiliation{Department of Physics, Centre for Quantum Coherence,
The Hong Kong Institute of Quantum Information Science and Technology,
and New Cornerstone Science Laboratory, The Chinese University of Hong Kong,
Shatin, New Territories, Hong Kong, China}

\author{Thierry~Chaneliere}
\affiliation{Universit\'e Grenoble Alpes, CNRS, Grenoble, France}

\author{Sylvain~Bertaina}
\affiliation{CNRS, Aix-Marseille Univ. University of Toulon, IM2NP, 13013, Marseille, France}

\author{Philippe~Goldner}
\affiliation{Chimie ParisTech, PSL University, CNRS, Institut de Recherche de Chimie Paris, 75005 Paris, France}

\author{Daniel~Est\`eve}
\author{Denis~Vion}
\author{Patrick~Abgrall}
\author{Patrice~Bertet}
\author{Emmanuel~Flurin}
\affiliation{Quantronics group, Service de Physique de l'\'Etat Condens\'e  (CNRS, UMR\ 3680),\\IRAMIS, CEA-Saclay, Universit\'e Paris-Saclay, 91191 Gif-sur-Yvette, France}

\email{emmanuel.flurin@cea.fr}

\date{\today}

\begin{abstract}

The ability to coherently control and read out qubits with long coherence times in a scalable system is a crucial requirement for any quantum processor. Nuclear spins in the solid state have shown great promise as long-lived qubits~\cite{steger_quantum_2012,pla_high-fidelity_2013, zhong_optically_2015}. Control and readout of individual nuclear spin qubit registers has made major progress in the recent years using individual electron spin ancilla qubits addressed either electrically~\cite{pla_coherent_2014,muhonen_storing_2014,madzik_precision_2022} or optically~\cite{maurer_room-temperature_2012,waldherr_quantum_2014,bradley_ten-qubit_2019,Abobeih2018,abobeih_fault-tolerant_2022}. Here, we present a new platform for quantum information processing, consisting of $^{183}$W nuclear spin qubits adjacent to an Er$^{3+}$ impurity in a CaWO$_4$ crystal, interfaced via a superconducting resonator and detected using a microwave photon counter at 10~mK. We study two nuclear spin qubits with $T_2^*$ of $0.8(2)~$s and $1.2(3)~$s and $T_2$ of $3.4(4)~$s and $4.4(6)~$ s, respectively. We demonstrate single-shot quantum non-demolition readout of each nuclear spin qubit using the Er$^{3+}$ spin as an ancilla. We introduce a new scheme for all-microwave single- and two-qubit gates, based on stimulated Raman driving of the coupled electron-nuclear spin system. We realize single- and two-qubit gates on a timescale of a few milliseconds, and prepare a decoherence-protected Bell state with a fidelity of 0.79 and $T_2^*$ of $1.7(2)~$s. Our results are a proof-of-principle demonstrating the potential of solid-state nuclear spin qubits as a promising platform for quantum information processing. With the potential to scale to tens or hundreds of qubits, this platform has prospects for the development of scalable quantum processors with long-lived qubits.

\end{abstract}

\maketitle

\begin{figure*}
    \centering
    \includegraphics[width=0.92\textwidth]{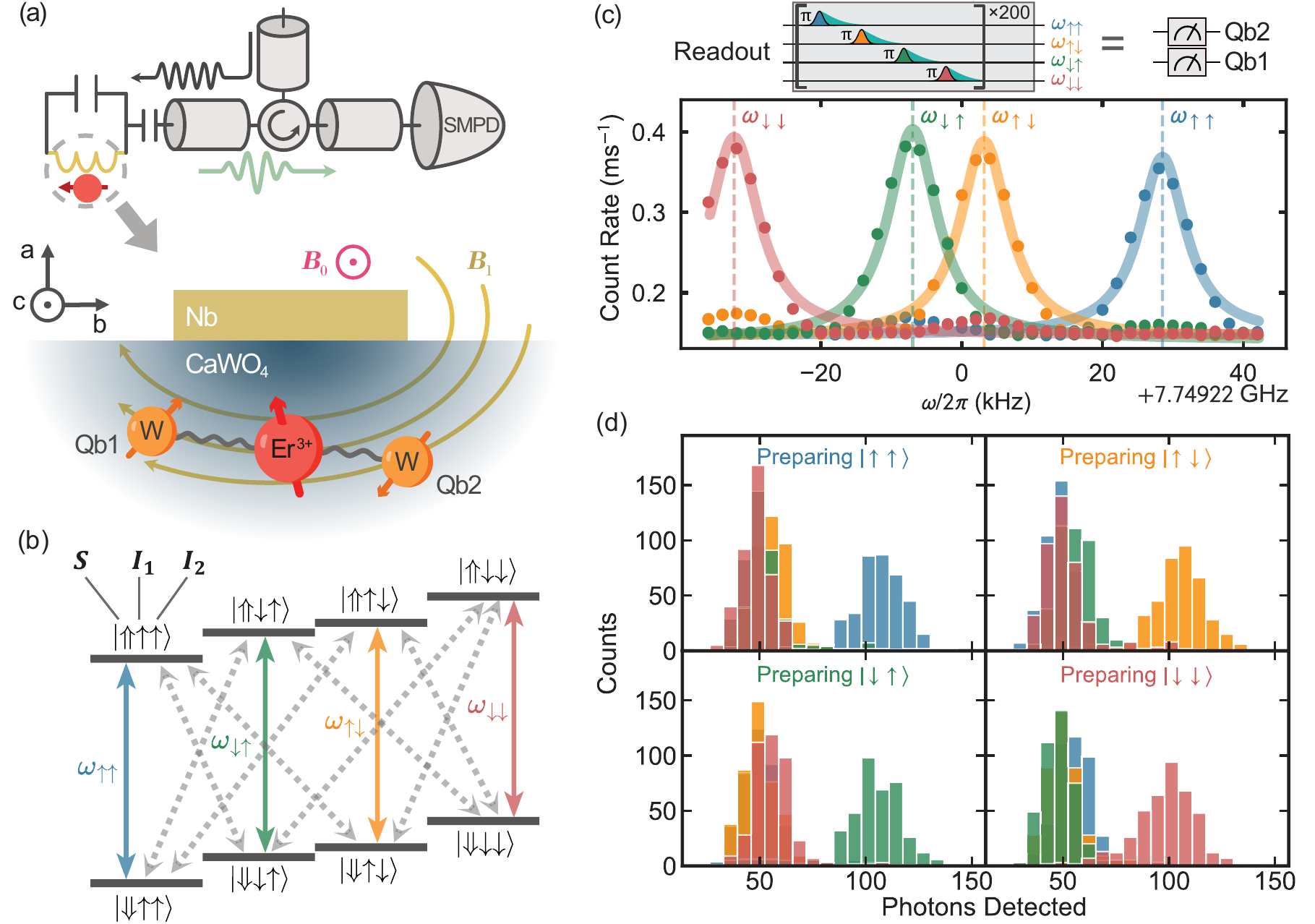}
    \caption{\textbf{Nuclear spin readout and preparation} (a) Top: An Er$^{3+}$ defect spin (red) is coupled to the inductor (yellow) of a superconducting lumped element resonator, which is capacitively coupled to an antenna. A circulator routes input pulses (black arrow) to the resonator and output signals (green arrow) to an SMPD. Bottom: cross section of the Nb nanowire inductor (yellow) on top of a CaWO$_4$ crystal (blue) with an Er$^{3+}$ spin (red) coupled to the magnetic field $B_1$ generated by the resonator. An external magnetic field $B_0$ is applied parallel to the nanowire. Two $^{183}$W nuclear spins (orange) are coupled to the Er$^{3+}$ spin. (b) Energy level diagram of the system of a spin $\mathbf{S}=1/2$ electron (double arrow) coupled to two spin $\mathbf{I_\mathrm{1,2}}=1/2$ nuclei (single arrows). Electron spin transitions occur at $\omega_{\uparrow\uparrow}, \omega_{\uparrow\downarrow}, \omega_{\downarrow\uparrow}$ and $\omega_{\downarrow\downarrow}$ (blue, orange, green and red, respectively) depending on the nuclear spin state. Simultaneous nuclear and electron spin-flip events are also possible via sideband transitions (grey dashed arrows). (c) Fluorescence of the electron spin as a function of excitation frequency when the $\ket{\uparrow\uparrow}$ (blue), $\ket{\uparrow\downarrow}$ (orange), $\ket{\downarrow\uparrow}$ (green) and $\ket{\downarrow\downarrow}$ (red) states are prepared.(d) Nuclear spin readout histograms. The $x$ axes correspond to the number of photons measured immediately after an excitation pulse at each frequency, $\omega_{\uparrow\uparrow}, \omega_{\uparrow\downarrow}, \omega_{\downarrow\uparrow}$ and $ \omega_{\downarrow\downarrow}$, is sent to the electron, summed over 200 repetitions at each frequency. This experiment was repeated $\sim$1000 times and the resulting number of photons detected was recorded each time to build up the histograms shown.}
    \label{fig:1}
\end{figure*}

\begin{figure*}
    \includegraphics[width=\textwidth]{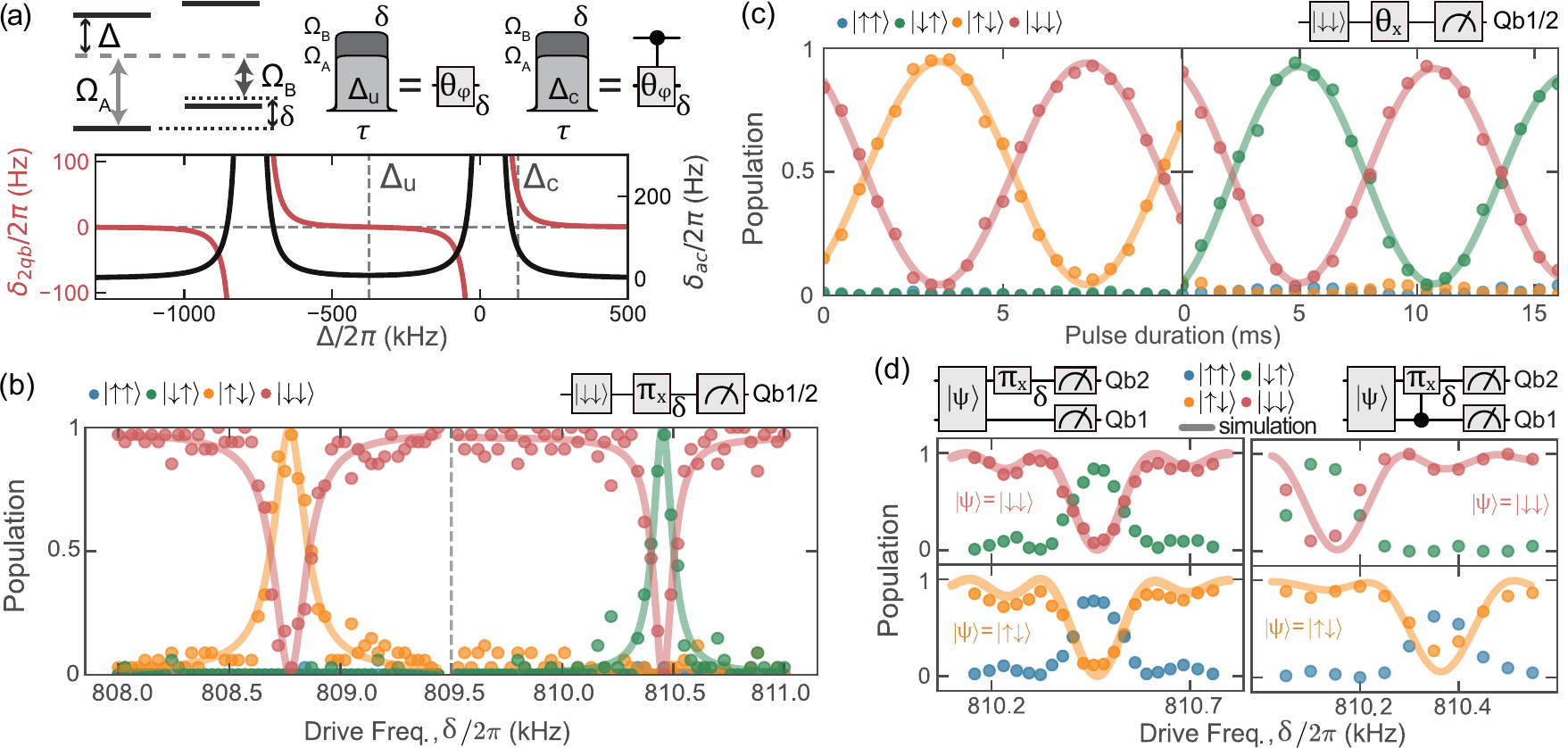}
    \caption{\textbf{Nuclear spin qubit spectroscopy and Rabi oscillations via stimulated Raman driving} (a) Top Left: energy levels of a single nuclear spin coupled to an electron spin and Raman drive fields $\Omega_A$ and $\Omega_B$ with frequency difference $\delta$. Top Right : Two-tone pulses used for unconditional and conditional driving, where $\theta$ is the rotation angle, $\phi$ is the rotation axis and the frequency difference $\delta$ is written next to the gate symbol only when swept. Bottom: Computed $\delta_{2qb}$ (red solid line) and $\delta_{ac}$ (black solid line) as a function of $\Delta$ for two drives with the same amplitude $\Omega /2\pi = 10$~kHz. Grey dashed line shows the unconditional driving condition $\Delta_u$, where $\delta_{2qb}=0$, and the conditional driving condition $\Delta_c$ used in this work. (b) Spectroscopy of the two nuclear spin transitions using unconditional Raman pulses $\Delta / 2\pi = -404.8$~kHz. Drive power below (above) 809.5~kHz was tuned to match the drive amplitude corresponding to a $\pi$-pulse for Qb1 (Qb2); continuous solid line is a Lorentzian fit. (c) Rabi oscillations of Qb1 (left) and Qb2 (right) as a function of pulse duration at 808.85~kHz and 810.45~kHz; drive amplitude was the same as that used in (b), continuous line is a sinusoidal fit. (d) Left (right): spectroscopy of Qb2 with $\Delta / 2\pi=-404.8$~kHz ($\Delta / 2\pi=110$~kHz). For the top figure, Qb1 was prepared in state $\downarrow$, while for the bottom figure Qb1 was prepared in state $\uparrow$. Solid line is the result of a simulation using the QuTiP package, based on the measured spin Hamiltonian parameters.}

    \label{fig:2}
\end{figure*}

\section{Introduction}

Nuclear spins in the solid-state are attractive candidates for quantum computing, offering long coherence times and high density. However, the typically weak coupling of nuclear spins to their environment that enables such long coherence times makes their control and readout challenging. Nuclear spin qubits are usually addressed through their hyperfine coupling to an electron spin ancilla, which then needs to be detected. Electrical detection of electron spin donors in silicon enabled the demonstration of long-lived single-$^{31}$P nuclear spin qubits in isotopically-enriched silicon~\cite{muhonen_storing_2014}, the operation of a two-$^{31}$P quantum gate~\cite{madzik_precision_2022}, and the control~\cite{fernandez_de_fuentes_navigating_2024} and preparation of a $^{123}$Sb $7/2$-nuclear spin in a Schr\"odinger-cat state~\cite{yu_creation_2024}. Electrical detection was used also for TbPc$_2$ molecular quantum dots, enabling the manipulation and readout of a $^{159}$Tb nuclear spin qubit~\cite{thiele_electrically_2014} and the implementation of Grover's search algorithm~\cite{godfrin_operating_2017}. Optical detection of individual NV center electron spins in diamond enabled the operation of a register of up to ten $^{13}$C nuclear spin qubits~\cite{bradley_ten-qubit_2019}, the fault-tolerant operation of a logical qubit built out of a $^{13}$C nuclear-spin qubit register~\cite{abobeih_fault-tolerant_2022}, and the demonstration of high-fidelity single- and two-qubit gates with a $^{14}\mathrm{N}$ nuclear spin~\cite{bartling_universal_2024}. Individual nuclear spin control has also been demonstrated using other color centres in diamond~\cite{beukers2024controlsolidstatenuclearspin, Smarak2022, Nguyen2019}, rare earth ions~\cite{Uysal2023} and SiC~\cite{babin2022}.

In this work, we present an alternative hybrid architecture to control and read out individual nuclear spin qubits via an electron spin ancilla by magnetically coupling this ancilla to a high-Q superconducting resonator. Requiring relatively straightforward fabrication techniques, this approach permits a single resonator to control numerous coherent spin defects. By detecting spin fluorescence~\cite{albertinale_detecting_2021} via a superconducting resonator using a superconducting circuit-based Single Microwave Photon Detector (SMPD)~\cite{lescanne_irreversible_2020,balembois_cyclically_2024}, detection of individual electron spins has recently been demonstrated \cite{wang_single-electron_2023}. Leveraging this technique, we use single Er$^{3+}$ spin impurities in a CaWO$_4$ crystal to control and read-out neighbouring spin-1/2 $^{183}$W nuclei via the hyperfine interaction. This method permits single-shot Quantun Non-Demolition (QND) readout of individual $^{183}$W nuclear spin qubits~\cite{travesedo_all-microwave_2024}. Individual all-microwave qubit control is achieved by driving a stimulated Raman transition. Two-qubit gates are achieved by taking advantage of induced AC Zeeman shifts on the energy level structure of the electron and nuclear spin system. We use this to construct a CNOT gate and generate Bell states of the two nuclei. We further show that certain Bell states exist in a so-called decoherence free subspace and are inherently resilient to magnetic field noise, extending $T_2^*$ from $0.8(2)~$s and $1.2(3)~$s for the individual nuclei to $1.7(2)~$s for the protected state.

\section{Device and Spin System}

We use a crystal of CaWO$_4$, which has tetragonal lattice symmetry, with a residual Er$^{3+}$ concentration of $\sim3.1$~ppb. At millikelvin temperatures, Er$^{3+}$:CaWO$_4$ behaves as an effective electron spin-1/2 system. This spin has a g-factor of 1.247 (8.38) parallel (perpendicular) to the $c$ axis of the crystal. Application of a static $B_0$ field lifts the two-state degeneracy via Zeeeman splitting. We pattern a planar niobium lumped-element superconducting resonator directly onto the substrate surface, as shown in Fig.\ref{fig:1}(a). The resonator has a small magnetic mode volume concentrated around a 300~nm wide, \SI{50}{\micro\meter} long inductor wire. This design maximises the magnetic field generated by the resonator around the wire and thus the inductive coupling $g_0$ to nearby electron spins. The resonator is placed inside a copper 3D cavity inside a dilution refrigerator and cooled to $\sim10$~mK. The resonator is coupled via a single antenna and circulator to an input port for drive fields and an output port to an SMPD. It is important to note that the methods described here are not unique to the nuclear or electron spin species chosen, the substrate crystal, or the resonator material, and are, in principle, extendable to a wide range of other spin species and materials. Further details of the experimental setup are given in~\cite{travesedo_all-microwave_2024}, in which the same sample and setup were used, and App.\ref{Appendix:experimental_setup}.


We apply a magnetic field $B_0$ of $446.8$~mT along the $c$ axis of the crystal. At this field, the resonator has a frequency of $7.7492$~GHz and a linewidth $\kappa/2\pi = 640~\mathrm{kHz}$. The substrate contains many Er$^{3+}$ spins with a distribution in transition frequency $\omega_S$ around a central bulk line. It is possible to probe a reduced subset of spins by tuning the magnetic field such that the resonator is detuned from the central line where the spectral density of the spins is greatly reduced and one can identify individual Er$^{3+}$ spins \cite{wang_single-electron_2023}. Using this method, we choose to focus on a single Er$^{3+}$ spin with interesting characteristics. The characteristic decay time of the spin was measured to be $0.8$~ms, dominated by spontaneous emission of photons into the resonator due to the Purcell effect~\cite{bienfait_controlling_2016}. We can use this to calculate the coupling of the Er$^{3+}$ spin to the resonator of $g_0/2\pi=5.6$~kHz (see App.\ref{Appendix:spectroscopy}).

The only isotope of W with non-zero nuclear spin, $^{183}$W, has spin-1/2 and a natural abundance of 14\%. The unit cell of CaWO$_4$ contains 10 W nuclei, meaning that it is likely that one or more spins in close proximity to the Er spin will have spin-1/2. The Er spin studied in this work was chosen due to its configuration of two neighbouring spin-1/2 W nuclei. The spin Hamiltonian of this system is as follows:

\begin{align}
\label{eq:spin_Hamiltonian}
\hat{H} = &\ \omega_S \hat{S}_{z} + \omega_{I,1} \hat{I}_{z,1} + \omega_{I,2} \hat{I}_{z,2}  \notag \\
          &+ A_\parallel^{(1)} \hat{S}_z \cdot \hat{I}_{z,1} + A_\perp^{(1)} \hat{S}_z \cdot \hat{I}_{x,1} \notag \\
          &+ A_\parallel^{(2)} \hat{S}_z \cdot \hat{I}_{z,2} + A_\perp^{(2)} \hat{S}_z \cdot \hat{I}_{x,2}
\end{align}

where $\omega_S$ is the Er spin frequency, $\hat{\mathbf{S}} = [\hat{S}_x, \hat{S}_y, \hat{S}_z]$ is the Erbium spin operator, $\omega_{I,i}$ are the nuclear spin frequencies, $\hat{\mathbf{I}}_i = [\hat{I}_{x,i}, \hat{I}_{y,i}, \hat{I}_{z,i}]$ are the nuclear spin operators and $A_\parallel^{(i)}$, $A_\perp^{(i)}$ are the $zz$ and $zx$ hyperfine coupling constants between the electron and nuclear spins, for $i\in1,2$. Here we have kept only the hyperfine terms proportional to $S_z$ according to the secular approximation. This Hamiltonian gives rise to an 8 level system, as shown in Fig.\ref{fig:1}(b). In this work, we utilise the two nuclear spins as qubits, designated Qb1 and Qb2, and the electron spin as an ancilla to control and read-out the nuclear spin qubits.

Transitions between the ground and excited state of the electron spin may be directly driven via the first term in Eq.\ref{eq:spin_Hamiltonian} by applying an AC magnetic field $\mathbf{B}_1(t)$ resonant with the electron spin transition via the superconducting resonator. The $zz$ coupling terms give rise to four distinct electron spin transition frequencies, $\omega_{\uparrow\uparrow}$, $\omega_{\downarrow\uparrow}$, $\omega_{\uparrow\downarrow}$ and $\omega_{\downarrow\downarrow}$, as indicated in Fig.\ref{fig:1}(b) by the four vertical arrows. The nuclei may in principle be driven in much the same way, however the resonator strongly suppresses fields at the requisite low drive frequencies of $\sim800$~kHz. Therefore, we choose to prepare the nuclear spin state via cross-transitions which simultaneously flip the electron and the nuclear spin~\cite{travesedo_all-microwave_2024}, as indicated by the grey arrows in Fig.\ref{fig:1}(b). These transitions are weakly allowed due to the presence of the $zx$ terms in the hyperfine Hamiltonian. We pump the system into the $\ket{\uparrow\uparrow}$ or the $\ket{\downarrow\downarrow}$ nuclear spin states by driving these cross transitions using a chirped pulse which covers all necessary frequencies and allowing the electron spin to subsequently decay~\cite{travesedo_all-microwave_2024}.

\section{Readout and Coherent Control of Single Nuclear Spins}

To detect the state of the nuclear spin qubits, we repeatedly excite the electron spin at each of the 4 possible frequencies, $\omega_{\uparrow\uparrow}$, $\omega_{\downarrow\uparrow}$, $\omega_{\uparrow\downarrow}$ and $\omega_{\downarrow\downarrow}$, sequentially and detect the photons emitted immediately after the excitation pulse. We sum the total number of photons counted over 200 repetitions at each frequency. The bandwidths of the resonator and SMPD, 640~kHz and $\sim200$~kHz, respectively, are sufficient to span all 4 frequencies simultaneously. Depending on the nuclear spin state, an excess of photon counts will be detected at one of the four frequencies. Detecting this excess allows us to perform QND readout  of the joint nuclear spin state. The number of counts detected as a function of electron spin drive frequency is shown in Fig.\ref{fig:1}(c) when the system is prepared in each of the four possible nuclear spin configurations (the data here have been corrected for slow drifts in the electron spin frequency, see App.\ref{Appendix:Frequency_drift} and~\cite{travesedo_all-microwave_2024} for more details). We see that the peak photon count rate occurs at a different frequency for each of the four nuclear spin states, and that the peaks are all clearly distinguishable, allowing for high fidelity readout of the nuclear spin state. We therefore fix the drive frequencies to $\omega_{\uparrow\uparrow}$, $\omega_{\downarrow\uparrow}$, $\omega_{\uparrow\downarrow}$ and $\omega_{\downarrow\downarrow}$ and measure the number of photons detected after preparing each of the 4 states for many repeated experiments to build the histograms shown in Fig.\ref{fig:1}(d). In each case, we see a clear separation between the number of photons detected after driving the transition corresponding to the prepared state vs the other three states. We threshold the number of photons detected to discriminate between the four states and measure probabilities of 0.91, 0.93, 0.92 and 0.95 to be in the desired state $\ket{\uparrow\uparrow}$, $\ket{\uparrow\downarrow}$, $\ket{\downarrow\uparrow}$ and $\ket{\downarrow\downarrow}$, respectively. These values are limited partially by state preparation errors and partially by cross-relaxation during readout via one of the sideband transitions. Cross-relaxation leads to a small probability of a nuclear spin flip event occurring during one of the 200 excitation pulses applied to the electron at each readout frequency during readout, see App.\ref{Appendix:discrimination} for further details.

Coherent control of individual nuclear spins has previously been achieved either by direct radio-frequency driving~\cite{thiele_electrically_2014,bradley_ten-qubit_2019}, or by coupled electron-nuclear dynamics relying on the repeated application of fast, broadband $\pi$ pulses to the electron spin at well-chosen time intervals~\cite{taminiau_detection_2012}. In our experiment, radio-frequency driving is not possible, and the dynamical decoupling approach is not well suited for a number of reasons (high magnetic field, cavity filtering of the excitation pulses, large electron spin relaxation rate). Instead, we use stimulated Raman transitions, which are largely used in atomic physics~\cite{leibfried_quantum_2003}. We show that they provide a new way to achieve all-microwave single- and two-nuclear-spin-qubit gates without populating the electron spin in its excited state. 

Stimulated Raman transitions are driven by applying two simultaneous microwave tones at different frequencies $\omega_S + A_\parallel^{(1)}/2 - A_\parallel^{(2)}/2 - \Delta $ and $\omega_S + A_\parallel^{(1)}/2 -A_\parallel^{(2)}/2 - \Delta - \delta$ (see Fig. \ref{fig:2}(a) and App.\ref{Appendix:Raman}), with amplitudes $\Omega_A$ and $\Omega_B$. Consider the coupling of the erbium spin to a single nuclear spin (case $A_\parallel^{(2)} = A_\perp^{(2)} = 0$ in Eq.1).
When $\delta$ is resonant with the frequency difference between the two ground-state levels $\delta_0 \equiv -\omega_{I,1} + A_\parallel^{(1)}/2 $, a transition between $\ket{\Downarrow \uparrow \uparrow}$ and $\ket{\Downarrow \downarrow \uparrow}$ can take place with an effective drive strength $\Omega_{\mathrm{Ram}} \approx \Omega_A \Omega_B A_\perp / 2 \omega_I [ 1 / \Delta - 1 / (\Delta - \omega_I) ]$ (see App.\ref{Appendix:Raman}). As long as $\Omega_{A,B} \ll \Delta$, the electron spin excited levels are never significantly populated, implying that the relatively short coherence time of the electron spin transitions plays no role in the system dynamics and therefore enabling high-fidelity nuclear-spin-qubit rotations. 

In addition, the Raman drives also cause ac-Zeeman frequency shifts of all the levels involved, due to the off-resonant driving of the electron transitions. As a result, the Raman resonance is found shifted by $\delta_{ac}$ from the expected frequency $\delta_0$ (see App.\ref{Appendix:Raman}). Moreover, when the second nuclear spin is considered ($A_\parallel^{(2)}, A_\perp^{(2)} \neq 0$), the value of $\delta_{ac}$ is in general conditioned on the state of the other spin, giving rise to two different values, $\delta^\uparrow_{ac}$ and $\delta^\downarrow_{ac}$. As such, we introduce $\delta_{2qb} \equiv \delta^\uparrow_{ac} - \delta^\downarrow_{ac}$, and $\delta_{ac} \equiv (\delta^\uparrow_{ac} + \delta^\downarrow_{ac})/2$. In the frame rotating at $\delta_0 + \delta_{ac}$, resonant stimulated Raman driving of nuclear spin 1 is described by the effective Hamiltonian $\Omega_{\mathrm{Ram}} I_{x,1} + [\omega_{I,2} - \delta_0 - \delta_{ac}] I_{z2} + \delta_{2q} I_{z1} I_{z2}$. The values of $\delta_{ac}$ and $\delta_{2qb}$, computed from an analytical model, are shown in Fig. \ref{fig:2}(a). Interestingly, there is a value of $\Delta$, which we call $\Delta_u \sim \omega_I /2$ (see Methods), for which $\delta_{2qb}=0$. When $\Delta = \Delta_u$, the Raman driving of one nuclear spin is therefore independent of the state of the other nuclear spin, a condition which is thus suitable for single-qubit gates. For other values of $\Delta$ where $\delta_{2qb} \neq 0$, the Raman resonance condition of one spin is dependent on the other spin state, providing a natural way to implement two-qubit gates, as described below.

We first perform Raman spectroscopy using the unconditional driving condition $\Delta/2\pi = -408.8$~kHz~$\approx \Delta_u/2\pi$, and $\Omega_A = \Omega_B$. The nuclear spin register is prepared in the $\ket{\downarrow\downarrow}$ state, after which a Raman pulse is applied whose detuning $\delta$ is swept, followed by nuclear spin readout. When the resonance condition is met, we observe a large change in the nuclear spin population to either the $\ket{\downarrow\uparrow}$ or the $\ket{\uparrow\downarrow}$ state (see Fig.~\ref{fig:2}(b)). The resonance linewidth is of order $100$~Hz or less, two orders of magnitude narrower than the electron spin linewidth, which confirms that the Raman driving occurs through virtual transitions of the electron excited states. The $\sim 2$~kHz frequency separation between the two ground-state nuclear spin transitions is due to the different values of $A_\parallel,A_\perp$ and $\omega_I$ for each $^{183}\mathrm{W}$ nuclear spin, due to their different position relative to the $\mathrm{Er}^{3+}$ ion (see App.\ref{Appendix:spectroscopy}). Knowing the required drive frequencies, we then prepare the $\ket{\downarrow\downarrow}$ state and sweep the drive duration, giving rise to coherent Rabi oscillations of each qubit, as shown in Fig.\ref{fig:2}(c). This allows us to calibrate $\pi$ pulses for Qb1 and Qb2, respectively.

We then study the spectrum of Qb2 after preparation of Qb1 in either the $\ket{\downarrow}$ or the $\ket{\uparrow}$ state (see Fig.~\ref{fig:2}(d)). If $\Delta \sim \Delta_u$ (left panel), the two resonances occur at the same frequency, confirming the unconditional driving condition. If $\Delta/2\pi  = \Delta_c /2\pi  = 110$\,kHz instead, the two resonances are shifted by $\delta_{2qb}/ 2\pi = 244(8)$~Hz. Both curves are quantitatively reproduced by simulations based on the spin Hamiltonian Eq.(1), using parameters determined by spectroscopic measurements (see App.\ref{Appendix:Raman}).

\begin{figure}[t]
    \includegraphics[width=0.44\textwidth]{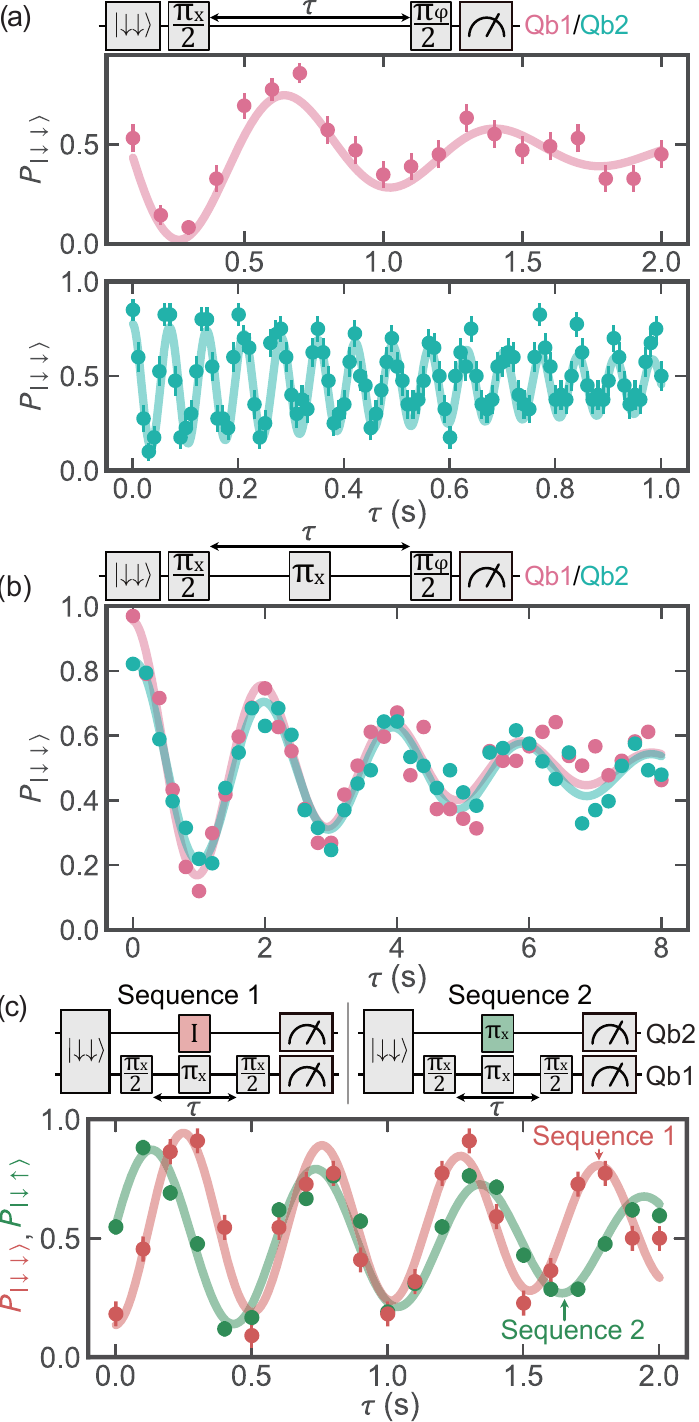}
    \caption{\textbf{Coherence of nuclear spin qubits}. All Raman drive pulses are applied in the unconditional regime, $\Delta = \Delta_u$. (a) Ramsey measurement of $T_2^*=0.8(2)$ s of Qb1 (top, pink) and $T_2^*=1.2(3)$ s of Qb2 (bottom, turquoise). (b) Hahn echo measurement of $T_2=3.4(4)$ s of Qb1 (pink) and $T_2=4.4(6)$ s of Qb2 (turqouise). (c) Spin-echo double resonance (SEDOR) measurement performed on Qb1, where sequence 1 and sequence 2 (as shown above) yielded the red and green curves respectively. We measure a nuclear-nuclear coupling rate $C_{zz}/2\pi = 0.8$~Hz.}
    \label{fig:3}
\end{figure}

We then proceed to measure the coherence characteristics of the nuclear spin qubits using unconditional Raman driving ($\Delta = \Delta_u$). We perform a Ramsey experiment on each qubit as shown in Fig.\ref{fig:3}(a) and extract $T_2^*$ times of 0.8(2)~s and 1.2(3)~s for Qb1 and Qb2, respectively. To our knowledge, these are the longest single-nuclear-spin Free-Induction-Decay times measured in a natural-abundance material by more than an order of magnitude. These times are on par with those measured in isotopically-enriched silicon~\cite{muhonen_storing_2014} and diamond~\cite{bartling_universal_2024}, which further confirms the interest of CaWO$_4$ for spin-based quantum computing~\cite{le_dantec_twenty-three-millisecond_nodate}. The residual decay is likely caused by a combination of $B_0$ drift and spectral diffusion due to nuclear spin bath reconfiguration. Coupled Cluster Expansion calculations predict a decay time of $0.2$~s, shorter than the measurements. We interpret the longer experimental $T_2^*$ as being due to partial polarisation of the $^{183}$W nuclear-spin bath by the pulses used to prepare the $\ket{\downarrow\downarrow}$ state, an effect which has been observed in similar systems~\cite{taminiau_universal_2014, madzik_controllable_2020}.

We performed Hahn echo measurements yielding $T_2$ times of 3.4(4)~s and 4.4(6)~s for qubits 1 and 2, respectively, as shown in Fig.\ref{fig:3}(b). Coupled Cluster Expansion calculations predict a Hahn echo decay time of $10$~s due to nuclear spin bath dynamics. We believe the extra decoherence is due to residual excitations of the Er$^{3+}$ ion, which occur at a rate $\bar{n} \Gamma_R$, $\bar{n} \approx 10^{-4}$ being the mean thermal occupation of the resonator mode, a value roughly consistent with the measured SMPD dark count rate. Energy relaxation of the two nuclear spin qubits likely occurs at a much lower rate, as indicated by ensemble measurements, which yield an upper bound of $\sim 10^{-6}~\mathrm{s}^{-1}$ for the relaxation rate of $^{183}\mathrm{W}$ proximal to Er$^{3+}$:CaWO$_4$~\cite{wang_month-long-lifetime_2024}, and was not directly measured.


\section{Two-qubit gate and Bell state characterization}
\label{Section:Bell_states}

Being in close proximity, the two nuclear spins are coupled by the dipolar interaction $C_{zz} I_{z1} I_{z2}$, and we determine $C_{zz}$ by performing a Spin Echo DOuble Resonance (SEDOR) experiment~\cite{abobeih_atomic-scale_2019}, as shown in Fig.\ref{fig:3}(c). This consists of a Hahn echo experiment performed on one nuclear spin qubit, the probe qubit, where a $\pi$ pulse is applied to the probe qubit and either no pulse (green) or a $\pi$ pulse (red) is applied to the other qubit at the middle of echo sequence. This yields a shift in frequency of the probe qubit equal to $C_{zz}/2$. We obtain $C_{zz} / 2\pi = 0.8$~Hz.


\begin{figure}
    \centering
    \includegraphics[width=0.45\textwidth]{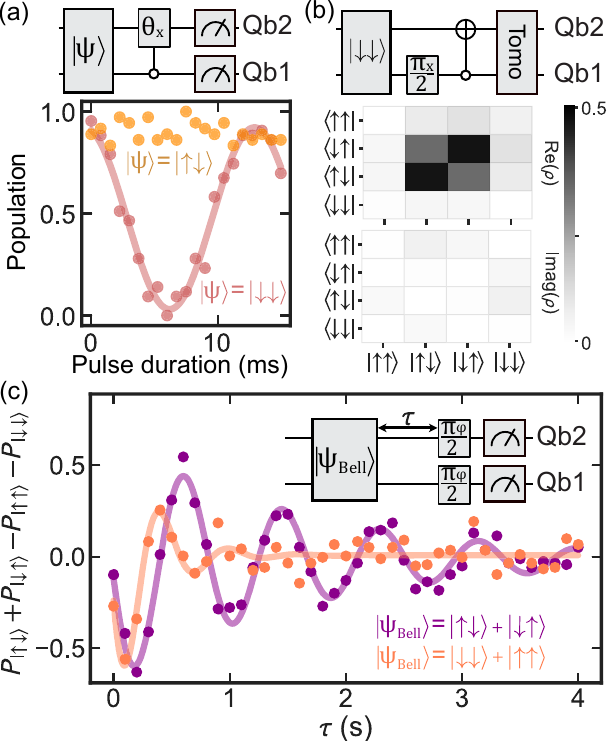}
    \caption{\textbf{Entanglement of two nuclear spins} (a) Top: Quantum circuit for driving Qb2 conditioned on the state of Qb1 - note that Qb2 is rotated only if Qb1 is in the $\ket{\downarrow}$ state, enacting an inverse CNOT gate when $\theta = \pi$. Botttom: Rabi oscillations when Raman driving Qb2 at 810.12~kHz when Qb1 is in the $\ket{\downarrow}$ (red circles) and $\ket{\uparrow}$ (orange circles) state. Red solid line is a cosine fit. (b) Top: Quantum circuit for preparation of the $\left(\ket{\downarrow\uparrow}+\ket{\uparrow\downarrow}\right)/\sqrt{2}$ state. A conditional $\pi$ pulse calibrated from the Rabi oscillations in (a) enacts an inverse CNOT gate. Details of tomography are given in App.\ref{appendix:state_tomography}. Bottom: density matrix reconstructed via state tomography of the $\left(\ket{\downarrow\uparrow}+\ket{\uparrow\downarrow}\right)/\sqrt{2}$ state. (c) Ramsey coherence measurement of the $\left(\ket{\downarrow\downarrow}+\ket{\uparrow\uparrow}\right)/\sqrt{2}$ (coral) state and the $\left(\ket{\downarrow\uparrow}+\ket{\uparrow\downarrow}\right)/\sqrt{2}$ (purple) state, yielding $T_2^*$ values of $0.3(1)$ and $1.7(2)~$s respectively. Inset: pulse sequence used for this experiment, where $\left(\ket{\downarrow\uparrow}+\ket{\uparrow\downarrow}\right)/\sqrt{2}$ Bell state preparation proceeds in the same manner as shown in (b) and $\left(\ket{\downarrow\downarrow}+\ket{\uparrow\uparrow}\right)/\sqrt{2}$ utilises an additional $\pi_\text{x}$ pulse on Qb2. }
    \label{fig:4}
\end{figure}


Such a low coupling rate is insufficient to implement any practical two-qubit entangling gate, and we instead use the conditional ac-Zeeman shifts of the Raman drive. We drive Qb2 using the $\Delta = \Delta_c$ conditional driving condition at 810.13~kHz when preparing the $\ket{\downarrow\downarrow}$ and $\ket{\uparrow\downarrow}$ states in Fig.\ref{fig:4}(a). We see that for a 6.72~ms drive pulse, the $\ket{\downarrow\downarrow}$ state is driven to the $\ket{\downarrow\uparrow}$ state, while the same pulse has no effect on the $\ket{\uparrow\downarrow}$ state. Thus, this pulse directly enacts an inverse CNOT gate.

We use this inverse CNOT gate to generate the Bell states $\left(\ket{\downarrow\downarrow}+\ket{\uparrow\uparrow}\right)/\sqrt{2}$ and $\left(\ket{\downarrow\uparrow}+\ket{\uparrow\downarrow}\right)/\sqrt{2}$. We perform tomography on these states and extract fidelities after state preparation and measurement error correction of 0.68 and 0.79 respectively; the reconstructed density matrix of the $\left(\ket{\downarrow\uparrow}+\ket{\uparrow\downarrow}\right)/\sqrt{2}$ state is shown in Fig.\ref{fig:4}(b), see App.\ref{appendix:state_tomography},\ref{Appendix:SPAM} for further details. In Fig.\ref{fig:4}(c), we measure the coherence of the $\left(\ket{\downarrow\downarrow}+\ket{\uparrow\uparrow}\right)/\sqrt{2}$ (coral) state and the $\left(\ket{\downarrow\uparrow}+\ket{\uparrow\downarrow}\right)/\sqrt{2}$ (purple) state, where the y axis is a linear combination of the populations of the four possible nuclear spin qubit states. This quantity measures the coherence of the Bell states. We see that the $\left(\ket{\downarrow\uparrow}+\ket{\uparrow\downarrow}\right)/\sqrt{2}$ state has a significantly longer $T_2^*$ than the $\left(\ket{\downarrow\downarrow}+\ket{\uparrow\uparrow}\right)/\sqrt{2}$ state; indeed it also exceeds that of either individual nuclear spin qubit. This is due to the fact that the $\left(\ket{\downarrow\uparrow}+\ket{\uparrow\downarrow}\right)/\sqrt{2}$ state is first-order insensitive to global magnetic field fluctuations, forming a so-called decoherence-free subspace~\cite{bartling_entanglement_2022}. The shorter lifetime compared to $T_2$ indicates a significant dephasing contribution from uncorrelated noise, likely caused by proximal nuclear spin reconfigurations. In contrast, the $\left(\ket{\downarrow\uparrow}+\ket{\uparrow\downarrow}\right)/\sqrt{2}$ state is sensitive to both correlated and uncorrelated magnetic noise, and thus displays a reduced coherence time, approximately half the $T_2^*$ of the individual qubits as expected.

Finally, we characterise the $\left(\ket{\downarrow\uparrow}+\ket{\uparrow\downarrow}\right)/\sqrt{2}$ state using state tomography and plot the resulting reconstructed density matrix in Fig.\ref{fig:4}(b). We extract a state fidelity (corrected for state preparation and measurement errors) of $0.79$ for the $\left(\ket{\downarrow\uparrow}+\ket{\uparrow\downarrow}\right)/\sqrt{2}$ state and $0.68$ for the $\left(\ket{\uparrow\uparrow}+\ket{\downarrow\downarrow}\right)/\sqrt{2}$ state (see App.\ref{appendix:state_tomography}). Simulations were used to estimate the contribution of different sources of error. When simulating generation and tomography of the $\left(\ket{\downarrow\uparrow}+\ket{\uparrow\downarrow}\right)/\sqrt{2}$ state including only coherent errors we find a final state fidelity of 0.978. When including incoherent errors the expected fidelity was 0.976. The discrepancy between simulated and experimental state fidelity may be accounted for by an increased thermal population of the Er spin during driving, see App.\ref{appendix:simulation} for further details.


\section{Conclusion}
In conclusion, we have presented a new platform for quantum information processing utilising nuclear spins in the solid state. We use an SMPD to detect the fluorescence of an Er$^{3+}$ spin coupled to a superconducting resonator, and use this to perform single-shot QND readout on two nearby nuclear spins. We utilise stimulated Raman drives to coherently control the nuclear spins and generate an entangled state. A key limitation of the experiment was measurement time: readout of the joint 2-qubit state requires $\sim 1$~s per shot. It should be possible to reduce the measurement time by orders of magnitude by increasing the Purcell-enhanced spin decay rate using a 3D resonator design to enhance resonator-spin coupling~\cite{Bienfait2016,Haikka2017}. To scale to larger system size, additional nuclear spin qubits could be interfaced using dynamical decoupling control techniques~\cite{bradley_ten-qubit_2019, Beukers2024}, or distant Er$^{3+}$ spins could be coupled using virtual photon exchange via the superconducting resonator~\cite{majer_coupling_2007, landig_virtual-photon-mediated_2019}. The resonator could also act as an interface to standard superconducting qubits. Beyond the Er:CaWO$_4$ system, many other crystal hosts and spin species could be investigated. The techniques demonstrated here are not limited to quantum information processing and may find utility in quantum sensing or electron paramagnetic resonance. The potential to explore a range of alternative systems and leverage the techniques described for new applications promises exciting new discoveries and developments in a platform that is largely unexplored.


\section*{Acknowledgements}
{We acknowledge technical support from P.~S\'enat, D. Duet, P.-F.~Orfila and S.~Delprat, and are grateful for fruitful discussions within the Quantronics group. We acknowledge support from the Agence Nationale de la Recherche (ANR) through the MIRESPIN (ANR-19-CE47-0011) project. We acknowledge support of the R\'egion Ile-de-France through the DIM QUANTIP, from the AIDAS virtual joint laboratory, and from the France 2030 plan under the ANR-22-PETQ-0003 grant. This project has received funding from the European Union Horizon 2020 research and innovation program under the project OpenSuperQ100+ and under the Marie Skłodowska-Curie grant agreement No 945298-ParisRegionFP, and from the European Research Council under the grant no. 101042315 (INGENIOUS). We thank the support of the CNRS research infrastructure INFRANALYTICS (FR 2054) and Initiative d'Excellence d’Aix-Marseille Université – A*MIDEX (AMX-22-RE-AB-199). We acknowledge IARPA and Lincoln Labs for providing the Josephson Traveling-Wave Parametric Amplifier. We acknowledge the crystal lattice visualization tool VESTA.}


\subsection*{Author contributions}
{The experiment was designed by J.T., J.O'S., E.F., and P.B. The crystal was grown by P.G. and characterized by EPR spectroscopy by S.B. The spin resonator chip was designed and fabricated by J.T. with the help of P.A. The SMPD was designed, fabricated, and characterized by L.P. under supervision of E.F. Data were acquired by J.O'S. and J.T. with the help of Z.W.H. and P.H. Data analysis and simulations were conducted by J.T., J.O'S., Z.W.H., and P.H. The manuscript was written by J.O'S., J.T. and P.B., with contributions from all co-authors. The project was supervised by P.B. and E.F.}

\newpage
\begin{appendix}

\section{Experimental setup and sample details}
\label{Appendix:experimental_setup}
\begin{figure}[b]
    \centering
    \includegraphics[width=0.75\columnwidth]{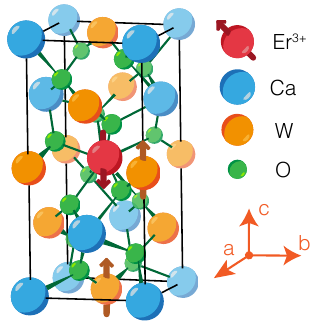}
    \caption{\textbf{Ball-and-stick diagram of the crystalline structure of CaWO$_4$.} An Er$^{3+}$ ion (red) substitutes a calcium (blue) atom. Tungsten (orange) has a 14.4\% probability of having a nuclear spin 1/2. The crystalline axis $(a, b, c)$ are shown in orange. The crystal has a tetragonal symmetry around the $c$-axis.}
    \label{fig:appendix_sample}
\end{figure}

CaWO$_4$ has a tetragonal crystalline structure, where an Er$^{3+}$ ion can substitute a calcium ion and enter the crystal as a paramagnetic impurity (see Fig. \ref{fig:appendix_sample}). The gyromagnetic tensor of Er$^{3+}$:CaWO$_4$ is diagonal the $(a, b, c)$ coordinate system. $\gamma_a=\gamma_b=\gamma_\perp= - 2\pi\cdot117.3$ GHz/T and $\gamma_c=\gamma_\parallel= -  2\pi\cdot17.45$ GHz/T \cite{antipin_aa_anisotropy_1981}.

The CaWO$_4$ crystal was grown at the Institut de Recherche de Chimie Paris and cut into a rectangular slab with dimensions 7~mm$\times$4~mm$\times$0.5~mm.  The surface of the sample is approximately parallel to the crystalline $(a, c)$ plane and the $c$ axis is parallel to the shorter edge.

A superconducting Nb resonator is fabricated on top of the surface of the sample. First, a 50~nm layer of Nb is sputtered on the polished surface of the sample. The resonator is patterned into a 30~nm  Al mask deposited on top of the Nb layer using electron beam lithography. The unmasked Nb is dry etched with a 1:2 mix of CF$_4$ and Ar. Finally, the Al mask is removed using MF-319.

\begin{figure*}[hbt!]
    \centering
    \includegraphics[width=0.95\textwidth]{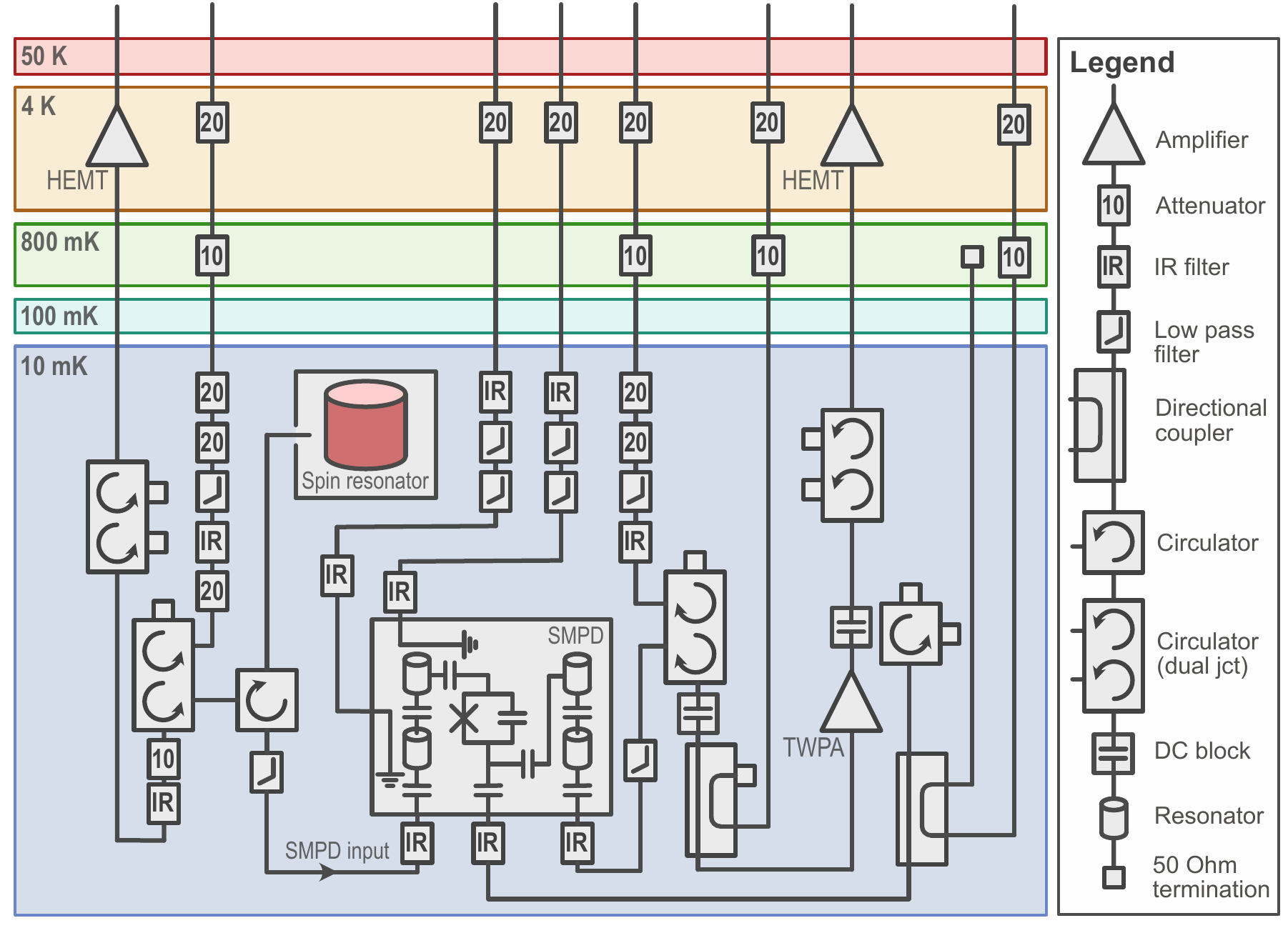}
    \caption{\textbf{Cryostat wiring diagram.} Schematic of the cryogenic microwave setup showing all attenuators and microwave components at all stages.}
    \label{fig:appendix_setup}
\end{figure*}

The resonator features a nanowire approximately parallel to the $c$-axis that creates the coupling field $\mathbf{B_1}$. The sample is then placed in a 3D cavity and coupled via an antenna to a coaxial cable. The output signal is routed through a circulator towards the Single Microwave Photon Detector (SMPD) input. A detailed wiring diagram is shown in Fig. \ref{fig:appendix_setup} .When the field $\mathbf{B_0}=446$ mT is applied approximately along the $c$-axis the resonator's frequency is 7.748 GHz, $Q_{\mathrm{ext}}$= $2.0 \cdot 10^4$ and $Q_{\mathrm{int}}$ = $2.8 \cdot 10^4$. This results in a linewidth of $\kappa/2\pi=640$~kHz. Additional details about the fabrication and characteristics of the resonator can be found in \cite{travesedo_all-microwave_2024}. 

\section{Electron and nuclear spin characterization}
\label{Appendix:spectroscopy}

\begin{figure}[t]
    \centering
    \includegraphics[width=\columnwidth]{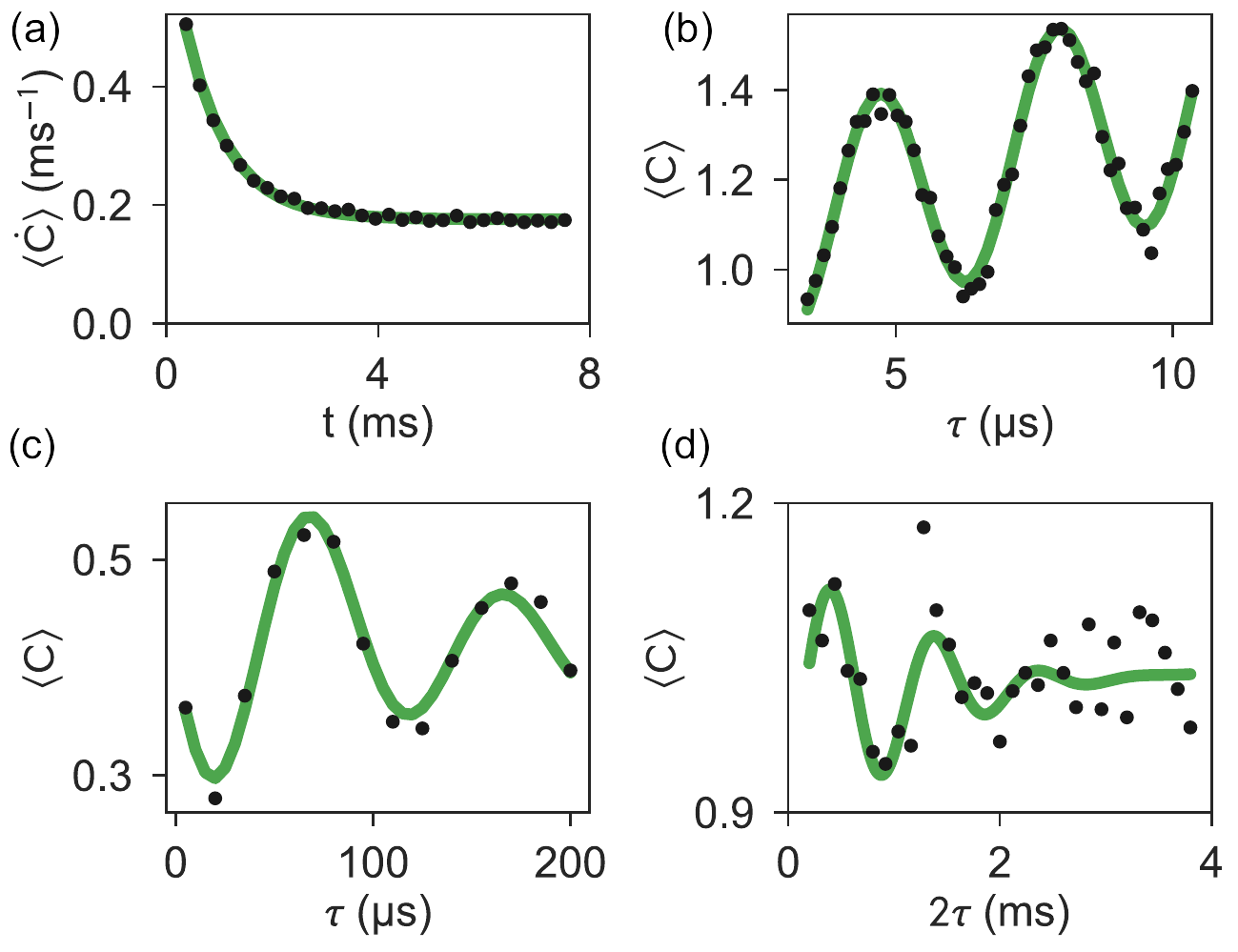}
    \caption{\textbf{Electron spin characterization.} Black dots represent measurements and green lines fits. (a) Fluorescence decay. Exponential fit yields $T_1$ = 0.8~ms (b) Rabi oscillations with a flattop pulse. Cosine fit yields $\Omega_e=/2\pi=310$~kHz (c) Ramsey oscillations and decay following nuclear spin state preparation. Decaying cosine fit yields $T_2^*=$\SI{170}{\micro\second}. (d) Echo decay with artificial oscillations following nuclear spin state preparation. Decaying cosine fit yields $T_2=$\SI{1.6}{\milli\second}.}
    \label{fig:appendix_electron}
\end{figure}

To detect a single Er$^{3+}$ spin, we excite the resonator with a Gaussian-shaped microwave pulse of \SI{5}{\micro\second} duration. If the frequency of the resonator and the spin are matched, the resonator drives the spin to the excited state; the spin subsequently relaxes via radiative emission due to the Purcell effect \cite{bienfait_controlling_2016}. We integrate the number of photons detected immediately after the spin excitation pulse during a time comparable to the $T_1$ of the spin which yields the ensemble averaged counts $\langle C \rangle$. Between the application of the pulse and the start of the photon counting, we wait \SI{120}{\micro\second} to avoid the extra counts created by the heating of the excitation pulse on the microwave lines. To perform spectroscopy, we measure $\langle C \rangle$ as a function of external magnetic field $\mathbf{B_0}$; an excess of photons indicates the presence of a spin. After finding the spin, we characterize $T_1$, $T_2^*$ and $T_2$ using fluorescence, Ramsey and spin echo experiments respectively. For the spin presented in this work we measure $T_1=0.8$~ms, $T_2^*=$\SI{170}{\micro\second} and $T_2=1.6$~ms (see Fig. \ref{fig:appendix_electron}). From the radiative decay rate $\Gamma_R = 1/T_1$, we calculate the coupling between the electron spin and the resonator $g_0/2\pi=5.6$~kHz using the expression for the Purcell rate when spin and resonator are on resonance, $\Gamma_R = 4g_0^2/\kappa$.

We characterize each nuclear spin by measuring $\omega_I$, $A_\parallel$ and $A_\perp$. To measure $A_\parallel$ we perform high-resolution spectroscopy on the electron spin with an \SI{80}{\micro\second} long Gaussian pulse as shown in Fig. \ref{fig:1}(c) of the main text. Preparing in each of the states will yield a different transition frequency for the electron spin. The separation between these frequencies is a direct measurement of $A_\parallel^{(1)}/ 2\pi=36(1)$~kHz and $A_\parallel^{(2)}/ 2\pi=19(1)$~kHz. Since the bandwidth of the resonator and the SMPD, 640~kHz and $\sim$200~kHz, are much larger than $A_\parallel^{(1)}$ and $A_\parallel^{(2)}$ we sweep the drive frequency of the pulse instead of $\mathbf{B_0}$. We leverage the high coherence of the nuclear spin to perform Hz resolution spectroscopy of the nuclear spin transitions. During the free-evolution time of the Ramsey sequence no drive tone is applied and the energy levels are not shifted. Therefore, the natural oscillation frequency is equal to $\delta_0=-\omega_I + A_\parallel/2$. From the data presented in Fig. \ref{fig:3}(a), we measure $-\omega_{I,1} + A_\parallel^{(1)}/2 = 2\pi \times 808.753(1)$~kHz and $-\omega_{I,2} + A_\parallel^{(2)}/2 = 2\pi \times 810.350(1)$~kHz for Qb1 and Qb2. 

To measure $A_\perp$ we compare the electron Rabi frequency to the nuclear Rabi frequency from the Raman driving. As detailed in section \ref{Appendix:Raman}, the frequency of the Rabi oscillations of a nuclear spin under a Raman drive is proportional to the hyperfine coupling $A_\perp$. From the data presented in Fig. \ref{fig:2}(d) we measure $\Omega_{\mathrm{Ram}}^{(1)} / 2\pi = 121(1)$~Hz and $\Omega_{\mathrm{Ram}}^{(2)} / 2\pi = 87(1)$~Hz.  From Eq. \ref{eq:raman_rabi} we obtain $A_\perp^{(1)} / 2\pi = 71(3)$~kHz and $A_\perp^{(2)} / 2\pi = 12.8(6)$~kHz. We note that there is a difference between the measured value of $A_\perp^{(1)}$ compared to the measurement using time trace statistics presented in \cite{travesedo_all-microwave_2024}. Table I compiles all the characterized values together.

\begin{table}[h!]
\centering
\begin{tabular}{|c|c|c|}
    \hline
                             & Qb1        & Qb2.    \\ \hline
    $\omega_I / 2\pi$ (kHz)  & -791(1)     & -801(1)  \\ \hline
    $A_\parallel/2\pi$ (kHz) & 36(1)      & 19(1)   \\ \hline
    $A_\perp/2\pi$ (kHz)     & 71(3)      & 12.8(6) \\ \hline
\end{tabular}
\label{tab:nuclear_spin}
\caption{Spectroscopically determined spin Hamiltonian parameters for Qb1 and Qb2.}
\end{table}

\section{Tracking electron spin frequency drifts}
\label{Appendix:Frequency_drift}

The electron spin experiences magnetic field fluctuations that manifest themselves as a drift of the transition frequency $\omega_S$. The origin of the fluctuations can be attributed to a combination of $B_0$ drift, charge noise, and nuclear spin bath reconfiguration. The drift is on the order of  $\sim5$~kHz, comparable to the linewidth of the electron-spin transitions, which potentially impacts nuclear spin readout. However, the drift is also slow, on the order of minutes, and it can be compensated for. Interleaved between measurements, we perform a drift correction algorithm to update the microwave drive frequencies. We run a Proportional-Integral (PI) loop that uses a Ramsey measurement to obtain the detuning between the old drive frequency of the $\ket{\Downarrow \uparrow \uparrow} \rightarrow \ket{\Uparrow \uparrow \uparrow}$ transition and the shifted value of $\omega_{\uparrow\uparrow}$. Then, all drive frequencies are corrected accordingly. A more in depth explanation of the PI-loop and the Ramsey measurement can be found in \cite{travesedo_all-microwave_2024}.

\section{Nuclear spin state discrimination and preparation}
\label{Appendix:discrimination}

\begin{figure}[t]
    \centering
    \includegraphics[width=0.85\columnwidth]{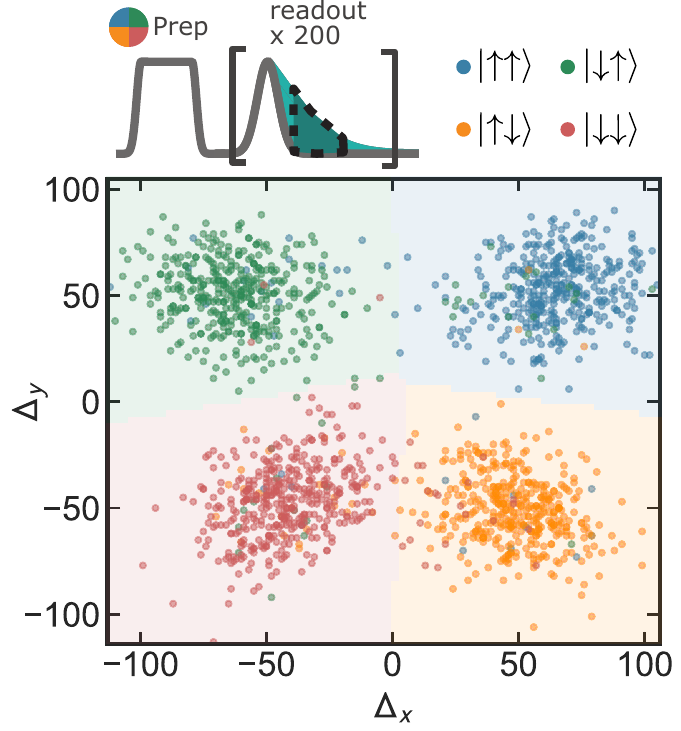}
    \caption{\textbf{Four-state readout of the nuclear spin state.}
    Each point corresponds to the value of $\Delta_x$ and $\Delta_y$ measured after preparing in a specific state. The colour indicates the prepared state: blue, orange, green and red correspond to $\ket{\uparrow \uparrow}$, $\ket{\uparrow \downarrow}$, $\ket{\downarrow \uparrow}$ and $\ket{\downarrow \downarrow}$ respectively. Background colour indicates the state assignment regions calculated using a $k$-means algorithm.}
    \label{fig:appendix_4state_readout}
\end{figure}

To measure the state of the nuclear spin we use the photon counting statistics obtained after sending a $\pi$-pulse at each of the resonant frequencies $\omega_{\uparrow\uparrow}$, $\omega_{\downarrow\uparrow}$, $\omega_{\uparrow\downarrow}$ and $\omega_{\downarrow\downarrow}$. We define $C_{\uparrow \uparrow}$, $C_{\downarrow \uparrow}$, $C_{\uparrow \downarrow}$ and $C_{\downarrow \downarrow}$ as the accumulated number of counts after applying a $\pi$-pulse with frequency matching each of the electron spin transitions respectively. We then introduce 

\begin{equation}
\begin{split}
   \Delta_x &= (C_{\uparrow \uparrow} + C_{\uparrow \downarrow}) - (C_{\downarrow \uparrow} + C_{\downarrow \downarrow}), \\
   \Delta_y &= (C_{\uparrow \uparrow} + C_{\downarrow \uparrow}) - (C_{\uparrow \downarrow} + C_{\downarrow \downarrow}).
\end{split} 
\end{equation}

\begin{figure*}[t]
    \centering
    \includegraphics[width=\textwidth]{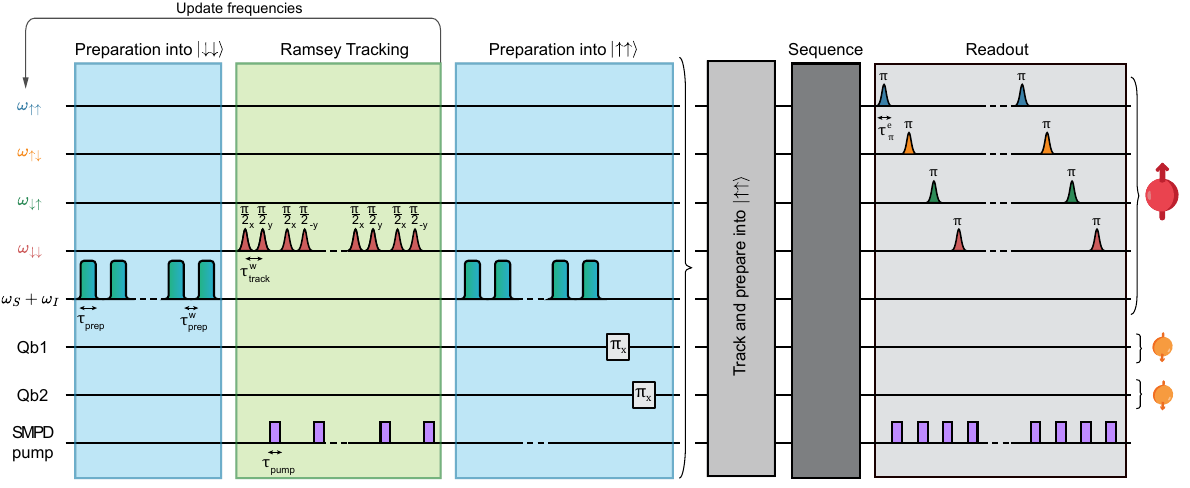}
    \caption{\textbf{Complete pulse sequence for preparation, tracking and readout.} For the preparation stage we apply a total of 50 chirped sideband pumping pulses. The pulses have a duration of $\tau_{\mathrm{prep}}=3.5$~ms and there is an interpulse delay of $\tau^\mathrm{w}_{\mathrm{prep}}=2$~ms for a total of $\sim$ 320 ms. The Ramsey tracking unit cell consists in two different Ramsey sequences repeated 10 times for a total of $\sim$ 20 ms. The Ramsey wait time, which limits the maximum range of the tracking, is $\tau^\mathrm{w}_{\mathrm{track}}$ = \SI{20}{\micro\second}. The readout primitive cell, which consists of four electron spin $\pi$ pulses of duration $\tau^e_\pi$ = \SI{80}{\micro\second} each followed by count integration, is repeated 200 times, adding up to $\sim$ 850 ms. For photon counting, an activation pulse is sent to the SMPD pump of duration $\tau_{\mathrm{int}}=1$~ms}
    \label{fig:sequence}
\end{figure*}
    
$\Delta_x$ ($\Delta_y$) effectively "traces out" the state of Qb2 (Qb1) as well as cancelling the contribution of the dark counts to the photon counting statistics. Figure \ref{fig:appendix_4state_readout} plots $\Delta_y$ versus $\Delta_x$ from the combined data presented in Fig. \ref{fig:1}(d). The colour of the points represents the state in which the system was prepared before the measurement took place. There are four well resolved distributions, each corresponding to a different prepared state of the two-nuclear-spin system. We use a $k$-means algorithm to separate the space corresponding to each distribution, displayed in the plot with the corresponding background colour. The boundaries approximately correspond to the $x=0$ and $y=0$ curves as expected. The skewness in the boundaries can be attributed to inaccurate calibration of the driving frequencies $\omega_{\uparrow\uparrow}$, $\omega_{\downarrow\uparrow}$, $\omega_{\uparrow\downarrow}$ and $\omega_{\downarrow\downarrow}$, which would result in different counting statistics for each state. The main limiting factor for this scheme is the probability for a state to cross-relax. When the electron spin is excited, there is a non-negligible probability for the electron spin to subsequently decay through an electron-nuclear transition with probability proportional to $\left(A_\perp / 2\omega_I \right)^2$, limiting the number of readout pulses that can be applied before the readout becomes non-QND. More details about the readout scheme can be found in \cite{travesedo_all-microwave_2024}

Preparation into arbitrary nuclear spin states is done through a combination of sideband pumping (also known as solid-effect Dynamic Nuclear Polarization \cite{abragam_a_procto_w_nouvelle_1958}) and single-qubit gates. Sideband pumping is achieved by driving the electron-nuclear transitions (represented as grey arrows in Fig. \ref{fig:1}(b) of the main text) with a $\tau_{\mathrm{prep}}=3.5$~ms length square pulse followed by $\tau^w_{\mathrm{prep}}$ = 2~ms of wait time, allowing the electron spin to decay \cite{travesedo_all-microwave_2024}. This sequence is then repeated 50 times to maximize the preparation efficiency. To prepare $\ket{\downarrow \downarrow}$ we drive the high frequency sidebands. To cover both Qb1 and Qb2 sidebands, the frequency of the pulse is chirped 50 kHz from $\omega_I / 2\pi$ + 760~kHz to $\omega_S / 2\pi$ + 810~kHz. To prepare other states, we apply the appropriate single-qubit gates after initializing in $\ket{\downarrow \downarrow}$.

An example of a full pulse sequence is displayed in Fig. \ref{fig:sequence}. It begins by preparing the nuclear spin state into $\ket{\downarrow \downarrow}$ via sideband pumping. The electron transition frequency is then tracked with a Ramsey sequence and the frequencies are updated accordingly. This is followed by another preparation stage, this time into any arbitrary state. In the case of Fig. \ref{fig:sequence}, the prepared state is $\ket{\uparrow \uparrow}$. We repeat the sideband pumping steps to avoid any cross-relaxation that might have taken place during the tracking stage. After an arbitrary sequence, the state of the nuclear spin is readout by interleaving electron $\pi$ pulses, with duration $\tau^e_\pi$ = \SI{80}{\micro\second}, at each of the possible transition frequencies and measuring the number of subsequent counts for an integration time $\tau_{\mathrm{int}}$  = \SI{1}{\milli\second}. In total the duration of the scheme, without considering the duration of the the arbitrary sequence is larger than 1 second. This low repetition time can be readily improved by increasing the coupling of the superconducting resonator to the electron spins, which in turn boosts the relaxation rate of the latter. Another avenue of improvement is optimization in the pumping scheme. Currently the duration and number of pumping pulses is large and a systematic study of these parameters will likely yield an enhancement in the preparation time.

\section{Stimulated Raman driving of nuclear spins}
\label{Appendix:Raman}

\begin{figure}[t]
    \centering
    \includegraphics[width=\columnwidth]{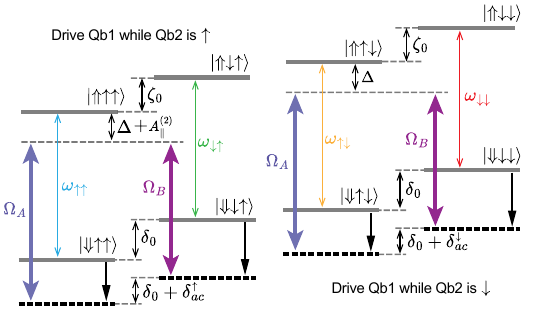}
    \caption{\textbf{Stimulated Raman driving scheme for an 8-level system.} Bare levels are represented as a grey continuous line. The purple thick double arrows show the two off-resonant drives with amplitudes $\Omega_A$ and $\Omega_B$. The drives have been represented twice to show the relevant quantities when driving Qb1 for the two states of Qb2. Black single arrows represent the ac-Zeeman shift from the off-resonant drive. Dashed black lines show the shifted levels due to the ac-Zeeman shifts.}
    \label{fig:SRS}
\end{figure}

In this section we derive the analytical expressions for the Rabi frequency of the Raman drive and the corresponding ac-Zeeman shifts for the Hamiltonian presented in Eq.~\ref{eq:spin_Hamiltonian}. The corresponding energy levels are shown in Fig.~\ref{fig:SRS}. In the sketched diagram, the system is under drive by two off-resonant tones in order to induce a stimulated Raman transition in Qb2. The two tones have amplitudes $\Omega_A$ and $\Omega_B$, defined with respect to the Rabi frequency of the electron at the same microwave amplitude $\Omega_{e, (A, B)}$:

\begin{figure}[t]
    \centering
    \includegraphics[width=\columnwidth]{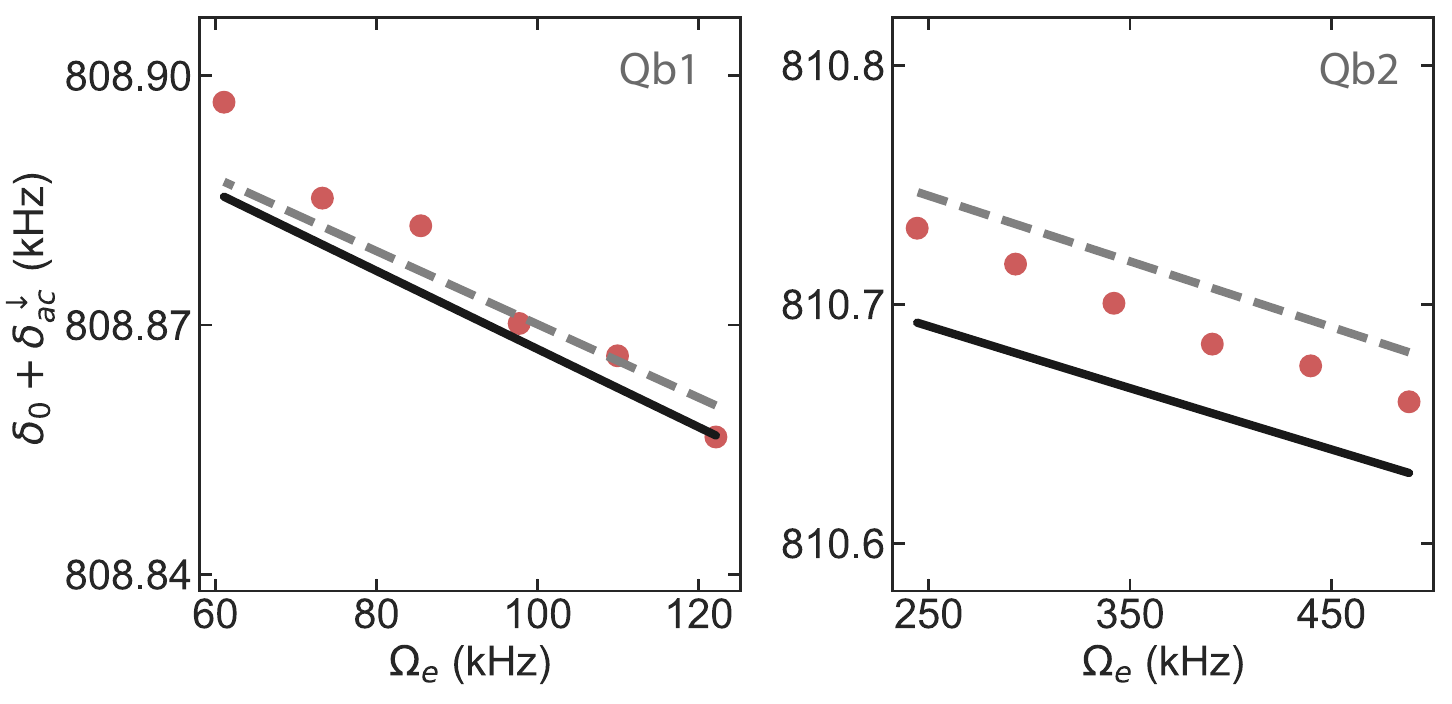}
    \caption{\textbf{ac-Zeeman shift as a function of drive amplitude.} Red dots are experimental data taken at $\Delta=1.4$~MHz and drive amplitudes $\Omega_A=\Omega_e=\Omega_B \cdot 1.57$. Dashed gray lines present the value obtained from Eq. \ref{eq:ac-zeeman}. Solid line is the result of a simulation using the QuTiP package, based on the measured spin Hamiltonian parameters.}
    \label{fig:ac-shifts}
\end{figure}

\begin{equation}
\begin{split}
    \Omega_{e, (A,B)} &= |\bar{\bar{\gamma}} \cdot \mathbf{B}_{1, (A,B)}| \cdot \bra{\Uparrow \uparrow \uparrow} S_x \ket{\Downarrow \uparrow \uparrow} \\
    &= |\bar{\bar{\gamma}} \cdot \mathbf{B}_{1}| / 2, \\
    \Omega_A &= \Omega_{e, A} \cdot \frac{1}{\sqrt{1 + 4\Delta^2/\kappa^2}}, \\
    \Omega_B &= \Omega_{e, B} \cdot \frac{1}{\sqrt{1 + 4(\Delta - \omega_I)^2/\kappa^2}}
\end{split}
\end{equation}

where $\mathbf{B}_{1,(A,B)}$ is the magnetic field at resonance generated by the nanowire under drive. The second term in the expressions for $\Omega_{A},~ \Omega_{B}$ originates from the filtering of the resonator, with linewidth $\kappa/2\pi = 640$~kHz. $\Omega_{e, (A, B)}$ is experimentally obtained by rescaling the Rabi frequency presented in Fig. \ref{fig:appendix_electron}(d) by the respective drive amplitudes. In Fig.~\ref{fig:SRS}, both tones are shown twice for each state of Qb1. Let us restrict ourselves to the subset of levels where Qb1 is in the $\uparrow$ state. When the frequency difference between the two drives is equal to $\delta_0 + \delta^\uparrow_{ac}$, the Qb2 is coherently driven from one state to the other following Rabi oscillations. Due to the presence of two excited states, two possible virtual paths need to be taken into consideration, the strength of each is given by

\begin{equation}
\begin{split}
    \Omega_{\mathrm{Ram},1} =& \frac{\Omega_A\Omega_B}{2\Delta} \cdot \frac{\bra{\Uparrow \uparrow \uparrow} S_x \ket{\Downarrow \uparrow \downarrow}}{\bra{\Uparrow \uparrow \uparrow} S_x \ket{\Downarrow \uparrow \uparrow}},  \\
    \Omega_{\mathrm{Ram},2} =& \frac{\Omega_A\Omega_B}{2\left(\Delta - \omega_I + A_\parallel^{(2)}/2\right)} \cdot \frac{\bra{\Uparrow \uparrow \downarrow} S_x \ket{\Downarrow \uparrow \uparrow}}{\bra{\Uparrow \uparrow \uparrow} S_x \ket{\Downarrow \uparrow \uparrow}}
\end{split}
\end{equation}

where the second term accounts for the driving of the sideband transition, which changes the corresponding matrix element. The addition of the two paths yields the total strength of the Raman drive

\begin{equation}
    \Omega_{\mathrm{Ram}} = \frac{A_\perp}{2\omega_I}\frac{\Omega_A\Omega_B}{2} \cdot \left(\frac{1}{\Delta} - \frac{1}{\Delta - \omega_I + A_\parallel^{(2)}/2}\right).
\label{eq:raman_rabi} 
\end{equation}

We now proceed to derive the ac-Zeeman shift $\delta_{ac}$. The presence of these two off resonant drives will inevitably induce ac-Zeeman shifts on the energy levels, which will in turn change the resonance condition for the spontaneous Raman transition from $\delta_0$ to $\delta_0 + \delta^\downarrow_{ac}$ ($\delta_0 + \delta^\uparrow_{ac}$) if Qb1 is in state $\downarrow$ ($\uparrow$). In the high field limit, where $\omega_I \gg A_\parallel^{(1)}, A_\parallel^{(2)}, A_\perp^{(1)}, A_\perp^{(2)}$ the shifts can be directly calculated as the combined shift of each drive on every electron transition,

\begin{equation}
\begin{split}
    \delta^\uparrow_{ac} &= \frac{\Omega_A^2}{4} \cdot \left(\frac{1}{\Delta} - \frac{1}{\Delta + \zeta_0 - \delta_0}\right) + \\
    &+ \frac{\Omega_B^2}{4} \cdot \left(\frac{1}{\Delta + \delta_0} - \frac{1}{\Delta + \zeta_0}\right), \\
    \\
    \delta^\downarrow_{ac} = \frac{\Omega_A^2}{4} \cdot &\left(\frac{1}{\Delta - A_\parallel^{(2)}} - \frac{1}{\Delta - A_\parallel^{(2)} + \zeta_0 - \delta_0}\right) + \\
    + \frac{\Omega_B^2}{4} \cdot &\left(\frac{1}{\Delta - A_\parallel^{(2)} +  \delta_0} - \frac{1}{\Delta - A_\parallel^{(2)} + \zeta_0}\right), \\
\end{split}
\label{eq:ac-zeeman}
\end{equation}

where $\delta_0=-\omega_I + A_\parallel^{(1)}/2$ and $\zeta_0=-\omega_I - A_\parallel^{(1)}/2$. We introduce the following quantities: $\delta_{ac} = \: (\delta^\uparrow_{ac} + \delta^\downarrow_{ac}) / 2$ and $\delta_{2qb} = \: \delta^\uparrow_{ac} - \delta^\downarrow_{ac}$. $\delta_{2qb}$ measures the relative ac-Zeeman shift on Qb2 depending on the state of Qb1. A single-qubit gate on Qb2 requires $\delta_{2qb}=0$ so that the drive is completely independent of the state of Qb1. This can be achieved by matching the two drive amplitudes $\Omega_A=\Omega_B=\Omega$ and setting the appropriate value of $\Delta=\Delta_u$. To calculate the latter we solve for $\delta_{2qb}=0$. At first order, the equation reduces to

\begin{equation}
\begin{split}
    \frac{\Delta_u\cdot(\Delta_u - A_\parallel^{(1)})\cdot(\Delta_u - A_\parallel^{(1)} + 2A_\parallel^{(2)})}{(\Delta_u - \omega_I)^2\cdot(\Delta_u + 2 A_\parallel^{(2)} - \omega_I)} = -1 
\end{split}
\end{equation}

This third order polynomial always has a real root. At zero-order ($\omega_I \gg A_\parallel^{(1)}, A_\parallel^{(2)}$) we obtain $\Delta_u = \omega_I/2$, which amounts approximately to $- 400$\,kHz as observed experimentally (Fig.2(d) of the main text).

For a two-qubit gate we try to maximize the value of $\delta_{2qb}$. We set $\Delta/2\pi=110$~kHz and $\Omega_B = 5\Omega_A$. This asymmetry is introduced to help preserve the Raman regime ($\Omega_A \ll \Delta$ and $\Omega_B \ll \Delta+\delta_0$) when the value of $\Delta$ is reduced. For this condition we measure $\delta_{2qb}/2\pi=244(8)$~Hz. 

Figure \ref{fig:ac-shifts} plots $\delta = \delta_0 + \delta^\downarrow_{ac}$ as a function of the drive amplitude at the following conditions: $\Delta=1.4$~MHz and $\Omega_A=\Omega_e=\Omega_B*1.57$. The analytic expressions as well as the simulations performed using the QuTiP package show quantitative agreement.

\section{State Tomography}
\label{appendix:state_tomography}

\begin{figure*}[t]
    \centering
    \includegraphics[width=0.9\textwidth]{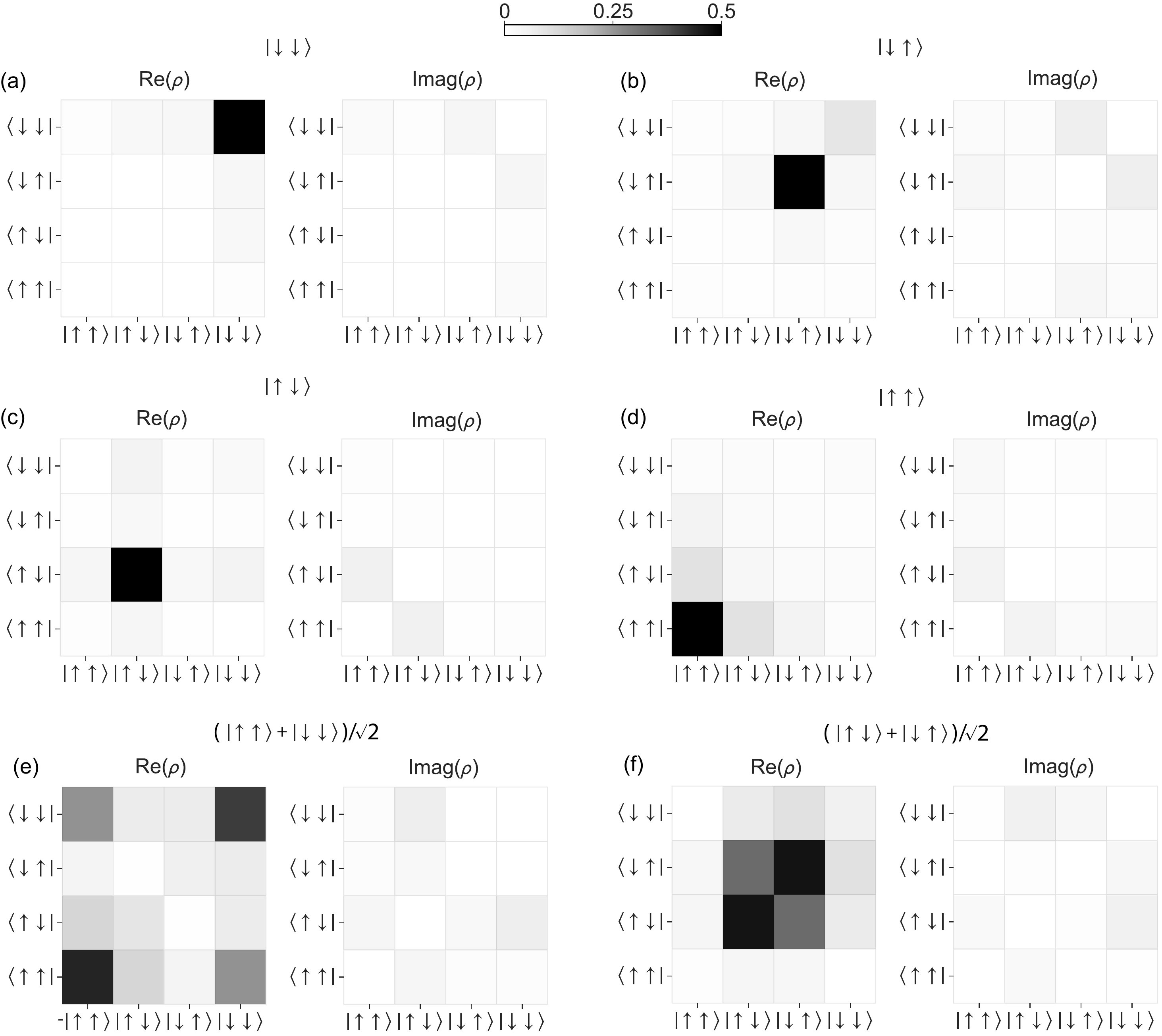}
    \caption{\textbf{State tomography of selected states.} Real (left) and imaginary (right) components of tomographically reconstructed density matrices of unentangled states (a,b,c,d) and Bell states (e,f). Extracted fidelities after SPAM error correction are 0.99, 0.96, 0.9, 0.93, 0.68 and 0.79 for the states shown in (a,b,c,d,e,f), respectively.}
    \label{fig:methods_tomo}
\end{figure*}

We define the fidelity of a measured state $\rho$ with respect to a target state $\rho_t$ as \cite{nielsen_quantum_2010}

\begin{equation}
    \mathcal{F} = \left(\text{Tr}\left\{\sqrt{\sqrt{\rho_t} \cdot \rho \cdot \sqrt{\rho_t}}\right\}\right)^2
\end{equation}

Figures \ref{fig:methods_tomo}(a,b,c,d) show the real and imaginary components of the tomographically reconstructed density matrices of the states $\ket{\downarrow \downarrow}$, $\ket{\uparrow \downarrow}$, $\ket{\downarrow \uparrow}$, $\ket{\uparrow \uparrow}$ respectively. The fidelity after correction for State Preparation And Measurement (SPAM) errors (see App.\ref{Appendix:SPAM} for details) for the single states is 0.99, 0.96, 0.9 and 0.93 respectively. $\ket{\downarrow \downarrow}$ has the highest fidelity compared to the other single states due to the state preparation protocol. As detailed in Appendix \ref{Appendix:discrimination}, this is the initial state we prepare into via sideband pumping. The other states are prepared through single-qubit gates starting from $\ket{\downarrow \downarrow}$.

Figures \ref{fig:methods_tomo}(e,f) show the real and imaginary components of the tomographically reconstructed density matrices of the Bell states $\left(\ket{\uparrow\uparrow}+\ket{\downarrow\downarrow}\right)/\sqrt{2}$ and $\left(\ket{\downarrow\uparrow}+\ket{\uparrow\downarrow}\right)/\sqrt{2}$, respectively. We measure fidelities for these states of 0.68 and 0.79 respectively. The cause of this infidelity was investigated with simulations and is discussed further in App.\ref{appendix:simulation}.

\section{SPAM errors}
\label{Appendix:SPAM}

SPAM errors arise from the non-ideality of the preparation and measurement schemes and limit the performance of any quantum computing platform. These errors can be characterized and subsequently removed from the results of complex measurements, such as state tomography. One such method is the $T$ matrix SPAM correction technique \cite{lidar_quantum_2013}. The $T$ matrix has a dimension of $2^n \times 2^n$, where $n$ is the size of the quantum register, in our case $n$=2. Each column of the $T$ matrix represents the probability distribution measured after preparing in a specific state. To remove SPAM errors, the inverse of the $T$ matrix is applied to the results of any measurement. From the data presented in Fig. \ref{fig:1}(d) we measure the following $T$ matrix:

\begin{equation}
    T = \begin{pmatrix}
0.91 & 0.02 & 0.05 & 0.01 \\
0.01 & 0.92 & 0 & 0.06 \\
0.05 & 0 & 0.91 & 0.03 \\
0 & 0.05 & 0.01 & 0.94 
\end{pmatrix} 
\end{equation}

SPAM error correction was performed for every tomography measurement and all fidelity values previously given account for this. Through SPAM correction we measure an increase of fidelity from 0.62 to 0.69 for the $\left(\ket{\uparrow\uparrow}+\ket{\downarrow\downarrow}\right)/\sqrt{2}$ state and an increase from 0.72 to 0.79 for the $\left(\ket{\downarrow\uparrow}+\ket{\uparrow\downarrow}\right)/\sqrt{2}$ state. Further study of the errors at play in the system will require the use of randomized benchmarking, which was not performed in the current experiment due to time constraints.

\section{Spin Dynamics Simulations and Error Analysis}
\label{appendix:simulation}

\begin{figure}[t]
    \centering
    \includegraphics[width=\columnwidth]{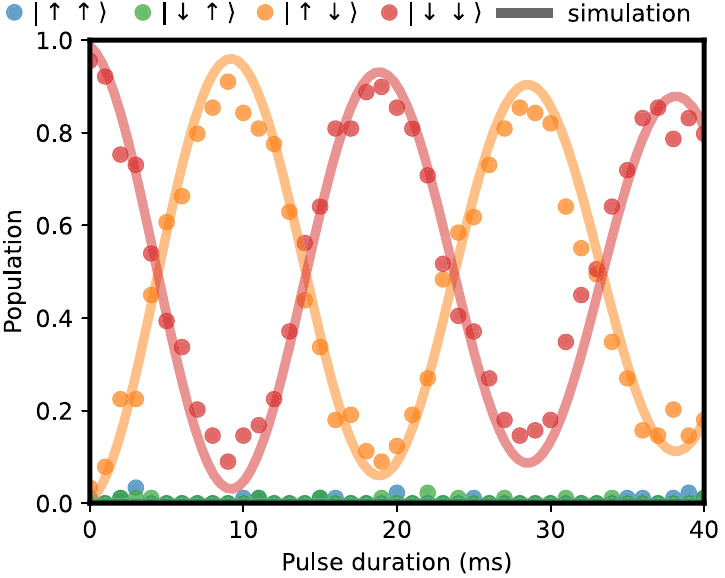}
    \caption{\textbf{Extended duration Raman Rabi on Qb1.} The decay time of the oscillations under driving $T_{2 \rho} = 150(35)$ ms was extracted from fitting. Solid line is a simulation with thermal population of the Er$^{3+}$ ancilla adjusted to $\bar{n}_{\text{th}} = 9.54\times10^{-3}$ to match the decay rate to the data.}
    \label{fig:Long_Rabi}
\end{figure}

We develop a simulation tool based on the QuTiP and Dynamiqs packages to simulate the system Hamiltonian under the influence of driving and $T_1$ and $T_2^*$ decay. The simulation considers an electron spin coupled to two nuclear spins and does not include a cavity or background spin bath. The spin system is described by
\begin{align}
\hat{H}_{0} = &\ \omega_S \hat{S}_{z} + \omega_{I,1} \hat{I}_{z,1} + \omega_{I,2} \hat{I}_{z,2}  \notag \\
          &+ A_\parallel^{(1)} \hat{S}_z \cdot \hat{I}_{z,1} + A_\perp^{(1)} \hat{S}_z \cdot \hat{I}_{x,1} \notag \\
          &+ A_\parallel^{(2)} \hat{S}_z \cdot \hat{I}_{z,2} + A_\perp^{(2)} \hat{S}_z \cdot \hat{I}_{x,2}
\end{align}
All parameters used in $\hat{H}_{0}$ are measured experimentally and detailed in Tab.\ref{tab:nuclear_spin}. An additional driving term with frequency $\omega$ is modeled by the Hamiltonian
\begin{equation}
    \hat{H}_d = \frac{\hbar\Omega_d}{2}(\hat{S}_{-}e^{i\omega t} + \hat{S}_{+}e^{-i\omega t}),
\end{equation}
where $\Omega_d$ is the pulse amplitude, $\hat{S}_{+}$ and $\hat{S}_{+}$ are the lowering and raising operators for the Er spin respectively. The pulse amplitude used in the simulation is calibrated using the measured electron spin Rabi frequency shown in Figure \ref{fig:appendix_electron}(b) and pulse durations are matched to those used in the experiment.

The simulation results follow from solving the master equation with the total time-dependent Hamiltonian
\begin{equation}
\frac{d\hat{\rho}}{dt} = \frac{1}{i \hbar} \left[ \hat{H}_{\text{tot}}, \hat{\rho} \right] + \sum_k \left( \hat{L}_k \hat{\rho} \hat{L}_k^\dagger - \frac{1}{2} \left\{ \hat{L}_k^\dagger \hat{L}_k, \hat{\rho} \right\} \right),
\end{equation}
where $\hat{H}_{\text{tot}} = \hat{H}_{\text{0}} + \hat{H}_{\text{d}}$. The electron energy relaxation process is modeled by the Lindblad operators $\hat{L}_{s_-} = \sqrt{\Gamma_{1} (\bar{n}_{\text{th}} + 1)} \, \hat{S}_{-}$ and $\hat{L}_{s_+} = \sqrt{\Gamma_{1} (\bar{n}_{\text{th}})} \, \hat{S}_{+}$, where $\Gamma_{1} = \frac{1}{T_1}$ is the electron relaxation rate. Dephasing of the nuclear spins is modeled by $\hat{L}_{n_\varphi,i} = \sqrt{\Gamma_{\varphi}} \, \hat{I}_{z,i}$ for $i=1,2$, where $\Gamma_{\varphi} = \frac{1}{T_2^*}$ is the nuclear dephasing rate. The parameter $\bar{n}_{\text{th}}$ models the finite thermal population of the electron spin. The effect of this term was observed experimentally on Qb1 as an increased dephasing rate under the influence of driving as shown in Fig.\ref{fig:Long_Rabi}. We observe Rabi oscillations decaying with a characteristic decay time $T_{2 \rho} = 150(35)$ ms, much faster than $T_2^*$ for this qubit. This may be attributed to an increase in $\bar{n}_{\text{th}}$ during driving which results in excess dephasing. A value of $\bar{n}_{\text{th}} = 9.54\times10^{-3}$ was used in the simulation in Fig.\ref{fig:Long_Rabi} to match the observed decay rate.

These simulations were used to investigate the origin of coherent and incoherent errors during Bell state preparation. We simulate the Bell state preparation sequence that yielded the $\left(\ket{\downarrow\uparrow}+\ket{\uparrow\downarrow}\right)/\sqrt{2}$ Bell state shown in Fig.\ref{fig:4}(b). We include subsequent basis rotations used for tomographic reconstruction and project the results into the $\{\ket{\uparrow},\ket{\downarrow}\}$ measurement basis to simulate the full experimental tomography process. We then perform maximum likelihood estimation on the resulting set of measurement results to reconstruct the density matrix and extract the fidelity to the target state. When including only coherent errors (i.e. removing all decay and dephasing effects) the simulation yields a fidelity of 0.978; when including incoherent errors in addition we find an expected fidelity of 0.976. The long coherence times of the two qubits ($\sim$ s) in comparison to the state preparation time ($\sim$ ms) explains the relatively small reduction in fidelity due to incoherent errors. However, the coherent errors are insufficient to account for the experimentally measured fidelity of 0.79. We attribute the remaining error to the finite thermal population of the Er electron spin under a driving field causing a reduced dephasing time as discussed previously and illustrated in Fig.\ref{fig:Long_Rabi}. Due to time constraints, measurement of $T_{2 \rho}$ under the exact conditions of the Bell state preparation sequence was not performed. However, simulations indicate that the same order of magnitude of $\bar{n}_{\text{th}}$ as estimated in Fig.\ref{fig:Long_Rabi} would be sufficient to account for the remaining error. Further investigation would be required to verify this.


\end{appendix}

\bibliography{NSQC}

\end{document}